\documentclass[preprint,aps,prd,floatfix,nofootinbib,11pt]{revtex4-2}
\pdfoutput=1
\usepackage{graphicx}
\usepackage{amsfonts}
\usepackage{amssymb}
\usepackage{xcolor}
\usepackage{natbib}
\usepackage{braket}

\usepackage{hyperref}

\usepackage{amsmath}
\usepackage{rotating}
\usepackage{amssymb,amsthm}
\usepackage{soul}

\usepackage{xcolor}

\usepackage{latexsym}
\usepackage{graphics}

\usepackage{amsfonts}
\usepackage{amsmath}
\usepackage{rotating}
\usepackage{amssymb}

\usepackage{amsmath}
\usepackage{rotating}
\usepackage{amssymb}
\usepackage{soul}

\usepackage{xcolor}

\usepackage{xcolor}
\def\beq{\begin{eqnarray}}  
	\def\eeq{\end{eqnarray}}

\def \bh {\mbox{{\bf h}}}

\begin{document}
	
	\title{Cosmological Solutions in Teleparallel $F(T,B)$ Gravity}
	
	\author{F. Gholami}
	\email{fateme.gholami@msvu.ca}
	\affiliation{Department of Mathematics and Statistics, Dalhousie University, Halifax, Nova Scotia, Canada, B3H 3J5}

	\author{A. Landry}
	\email{a.landry@dal.ca}
	\affiliation{Department of Mathematics and Statistics, Dalhousie University, Halifax, Nova Scotia, Canada, B3H 3J5}


\begin{abstract}
In this paper, we find several teleparallel $F(T,B)$ solutions for a Robertson--Walker (TRW) cosmological spacetime. We first set and solve the $F(T,B)$-type field equations for a linear perfect fluid. Using similar techniques, we then find new $F(T,B)$ solutions for non-linear perfect fluids with a weak quadratic correction term to the linear equation of state (EoS). Finally, we solve for new classes of $F(T,B)$ solutions for a scalar field source by assuming a power-law scalar field and then an exponential scalar field in terms of the time coordinate. For flat cosmological cases ($k=0$ cases), we find new exact and approximate $F(T,B)$ solutions. For non-flat cases ($k=\pm 1$ cases), we only find new teleparallel $F(T,B)$ solutions for some specific and well-defined cosmological expansion subcases. We conclude by briefly discussing the impact of these new teleparallel solutions on cosmological processes such as dark energy (DE) quintessence and phantom   {energy models}.
\end{abstract}

\maketitle



\newpage

\section{Introduction}\label{sect1}

{Teleparallel theories of gravity are an important class of alternative theories where all quantities and symmetries are defined in terms of the coframe and the spin- connection~\cite{Krssak:2018ywd,Krssak_Saridakis2015,Coley:2019zld,HJKP2018,HJKP2018a,Bahamonde:2021gfp,Cai_2015,preprint,SSpaper,TdSpaper,coleylandrygholami,nonvacSSpaper,nonvacKSpaper,roberthudsonSSpaper}}. The most appropriate definition of symmetry in this theory is an affine symmetry. In a teleparallel geometry, an affine frame symmetry on the frame bundle is defined for a coframe/spin-connection pair and  by a {field} ${\bf X}$ that satisfies 
\cite{Coley:2019zld,preprint,coleylandrygholami}
\begin{equation}
	\mathcal{L}_{{\bf X}} \bh_a = \lambda_a^{~b} \,\bh_b \text{and } \mathcal{L}_{{\bf X}} \omega^a_{~bc} = 0, \label{Intro:FS2}
\end{equation}
where $\omega^a_{~bc}$ is the spin-connection defined with respect to the frame $\bh_a$ by using the Cartan--Karlhede algorithm, and $\lambda_a^{~b}$ is the linear isotropy group component. This definition is the {\it {frame-dependent}} analogue of the symmetry definition as presented in refs.~\cite{HJKP2018,HJKP2018a}.

Using this algorithm, the invariant coframes and the corresponding spin-connection respecting the imposed affine frame symmetries can be constructed. However, there are some cosmological teleparallel spacetime geometries which are invariant under the full $G_6$ Lie algebra of affine symmetries~\cite{preprint,coleylandrygholami}. In addition, the proper coframes were obtained in each case, and the pure symmetric field equations (FEs) were found. There are the teleparallel Robertson--Walker (TRW) geometries~\cite{preprint,coleylandrygholami} with a Robertson--Walker (RW) metric $g_{\mu\nu}$ form and a constant parameter $k = (-1,0,1)$~\cite{preprint,coleylandrygholami}. This parameter $k$ is often interpreted as the constant spatial curvature in the RW pseudo-Riemannian metric, but the Riemann tensor is zero in teleparallel spacetimes. The parameter $k$ is often considered as a three-dimensional space curvature in the metric-based approach, but it is a part of the torsion scalar under a four-dimensional teleparallel spacetime. In TRW geometries, an appropriate coframe/spin-connection pair leads to the full trivial (null) antisymmetric part of the FEs. There are some TRW geometries where the frame and/or corresponding spin-connection admits the full $G_6$ Lie algebra defined by its six Killing Vectors (KVs). In the literature, there are, for   {non-zero}~$k$, some investigated geometries which do not yield a $G_6$ Lie algebra. For $k=\pm 1$ cases in the literature, there are some solutions involving the use of inappropriate tetrads and/or spin-connections. Recently, new coframe/spin-connection pairs satisfying a $G_6$ symmetry group have been constructed~\cite{Hohmann:2015pva,HJKP2018,HJKP2018a,Hohmann:2018rwf}. The most recent of these achievements leading to new teleparallel solutions can be found in refs.~\cite{Coley:2019zld,coleylandrygholami}.

A lot of works across the literature have been conducted with the analysis of flat ($k=0$) TRW cosmological models (see refs.~\cite{Bahamonde:2021gfp,Cai_2015} and references therein). In particular, specific forms for $F(T)$ have been investigated by using specific ansatz, and reconstruction methods have been explored extensively (where the function $F(T)$ is reconstructed from assumptions on the models). Dynamical system methods (e.g., fixed-point and stability analyses) in flat TRW models have been widely used in refs.~\cite{Bahamonde:2021gfp,coley03,BahamondeBohmer,Kofinas,BohmerJensko,aldrovandi2003}, including the study of stability conditions of the standard de Sitter fixed point. The non-zero $k$ geometries were recently studied in bounce and inflation models~\cite{bounce,Capozz} (i.e., the analysis is only applicable for the   {$k=1$ case).} Finally, perturbations were also studied in non-flat cosmology~\cite{inflat}. In a recent paper, we found as exact $k=0$ solutions of the FEs a combination of two power-law terms with the cosmological constant~\cite{coleylandrygholami}. For the $k=-1$ and $+1$ cases, the differential equation was linearized and then, by using the $k=0$ exact solution (from the dominating term) and a linear correction term, a rigorous stability test was performed to determine specific conditions depending on power-law ansatz, EoS, and cosmological parameters for stable TRW solutions and models~\cite{coleylandrygholami}. Even after the development of $F(T)$ TRW geometries and solutions, it is very possible to go further with the same approach.

After studying $F(T)$ TRW geometries and the perfect fluid solutions, it is necessary to generalize by taking into account the Ricci curvature scalar $\overset{\ \circ}{R}$. Accessing the $\overset{\ \circ}{R}$ influence on teleparallel spacetime geometry is possible by using a boundary variable defined as   {$B=\overset{\ \circ}{R}+T$.} There have been in recent years some studies concerning $F(T,B)$ cosmology and its {  {solutions \cite{leonftbcosmo1,leonftbcosmo2,leon2,leon3,cosmobaha,dixit,cosmoftbmodels,cosmoftbenergy,ftbcosmo2,ftbcosmo3,ftbcosmo4,ftbcosmo5,ftbcosmo6}.} Some of the studies are interesting with regard to cosmological models, such as bouncing, dark energy (DE) quintessence, and phantom energy (or negative energy), to name only a few~\cite{cosmobaha,cosmoftbmodels,ftbcosmo2,ftbcosmo3,ftbcosmo4,ftbcosmo5,ftbcosmo6}. The last physical processes are of great interest in the literature for a number of mathematical approaches, not only for teleparallel $F(T)$- and $F(T,B)$-type gravity.} In particular, there have been some very specific ansatzes used on power-law solutions containing a scalar field source. At the same time, there have been studies of the stability and convergence of solutions. There are even variants to the generalizations of teleparallel $F(T)$ gravity solutions such as $F(T,\phi)$, $F(T,T_G)$, and so on~\cite{Kofinas,ftphicosmo}. These last intermediate cases are just other, different ways of generalizing teleparallel gravity theories. For a typical cosmological spacetime, the most direct and straightforward of the TRW spacetime geometries is the generalization to the teleparallel $F(T,B)$-type theory~\cite{Bahamonde:2021gfp,leonftbcosmo1}. However, the studies mentioned above do not actually lead to $F(T,B)$ solutions other than combinations of power-law in $T$ and $B$. There is room to further develop the $F(T,B)$ exact solutions not only for scalar field sources but also for perfect linear and non-linear fluid cases. Solving in detail the $F(T,B)$ TRW FEs  {by using the most appropriate ansatz} is the fundamental aim of this paper. Several teleparallel $F(T,B)$ solutions are in principle possible and will generalize $F(T)$-type TRW spacetime solutions. These new $F(T,B)$ solutions will take into account $T$ as in $F(T)$ and also the curvature $\overset{\ \circ}{R}$ of the affine Levi--Civita connection by using the boundary variable $B$.

This paper aims to find exact and approximate $F(T,B)$ solutions for perfect fluids with linear and non-linear equations of state (EoSs) and for scalar fields with a potential $V(\phi)$. We will first proceed in Section~\ref{sect2} to a summary of Teleparallel $F(T,B)$ gravity and present the coframe and spin-connection components for the TRW geometry. In Section~\ref{sect3}, we will set the $k=0$, $-1$ and $+1$ general FE systems to solve in this paper. In Sections~\ref{sect4} and~\ref{sect5}, we will solve the FEs for linear and quadratic perfect fluids, exact and approximate solutions, respectively. In Section~\ref{sect6}, we will do the same derivations with some scalar field sources. For the rest, we will conclude in Section~\ref{sect7} with a summary of the new teleparallel $F(T,B)$ solutions, their nature and future possible developments on cosmologically more complete and relevant models. This is a study on the main possible teleparallel $F(T,B)$ cosmological solutions and some implications.


\section{Summary of Teleparallel \boldmath{$F(T,B)$} Gravity and FLRW Geometry}\label{sect2}

\subsection{Teleparallel $F(T,B)$ Gravity Field Equations}\label{sect21}

\noindent The action integral of teleparallel $F(T,B)$ gravity with matter (or scalar field) is~\cite{Bahamonde:2021gfp,cosmobaha,dixit,cosmoftbmodels,cosmoftbenergy,ftbcosmo2,ftbcosmo3,ftbcosmo4,ftbcosmo5,ftbcosmo6}:
\begin{align}\label{201}
	S_{F(T,B)}=\int\,d^4x\,\left[h\,\frac{F(T,B)}{2\kappa}+\mathcal{L}_{matter}\right], 
\end{align}
where $h$ is the coframe determinant and $\kappa$ is the coupling constant and $B=\overset{\ \circ}{R}+T$. From the least-action principle, the symmetric and antisymmetric parts of FEs for any teleparallel $F(T,B)$ theory {are}~\cite{ljsaidftb}: 
\small
	\begin{eqnarray}
		\kappa\,\Theta_{\left(ab\right)} &=& F_T \overset{\ \circ}{G}_{ab}+S_{\left(ab\right)}^{\;\;\;\;\mu}\,\left(F_{TT}\partial_{\mu} T+F_{BB}\partial_{\mu} B\right)+\frac{g_{ab}}{2}\,\left[F-T\,F_T-B\,F_B\right]+\left(\partial_a\partial_b-g_{ab}\,\partial_{\sigma}\partial^{\sigma}\right)\,F_B,
		\nonumber\\
		\label{202a}
		\\
		0 &=& S_{\left[ab\right]}^{\;\;\;\;\mu}\,\left(F_{TT}\partial_{\mu} T+F_{BB}\partial_{\mu} B\right), \label{202b}
	\end{eqnarray}
\normalsize
where $\overset{\ \circ}{G}_{ab}$ is the Einstein tensor, $S_{ab}^{\;\;\;\;\mu}$ is the superpotential, $g_{ab}$ is the gauge metric and $\Theta_{\left(ab\right)}$ is the energy-momentum.

\subsection{Energy-Momentum, Fluid and Scalar Field Parameters}\label{sect22}

\noindent The canonical energy-momentum is defined from $\mathcal{L}_{Matter}$ of Equation~\eqref{201} as~\cite{Bahamonde:2021gfp}:
\begin{align}\label{203a}
	\Theta_a^{\;\;\mu}=\frac{1}{h} \frac{\delta \mathcal{L}_{Matter}}{\delta h^a_{\;\;\mu}}.
\end{align}

{The antisymmetric and symmetric parts of Equation~\eqref{203a} are, respectively~\cite{Bahamonde:2021gfp,SSpaper,nonvacSSpaper,nonvacKSpaper,roberthudsonSSpaper,scalarfieldKS,scalarfieldTRW}}:
\begin{equation}\label{203b}
	\Theta_{[ab]}=0,\qquad \Theta_{(ab)}= T_{ab},
\end{equation}

With the definitions of $\Theta_a^{\;\;\mu}$ in Equations~\eqref{203a} and~\eqref{203b}, we find that the energy-momentum conservation law is~\cite{Bahamonde:2021gfp,SSpaper,nonvacSSpaper,nonvacKSpaper,roberthudsonSSpaper,scalarfieldKS,scalarfieldTRW}:
\begin{align}\label{203}
	{\overset{\ \circ}{\nabla}}\,_{\nu}\,\Theta^{\mu\nu} = 0.
\end{align}

Equation~\eqref{203} is essentially the General Relativity (GR) conservation law for the case where the hypermomentum is ${\mathfrak{T}^{~\mu\nu}_{\sigma}=0}$~\cite{nonvacSSpaper,nonvacKSpaper,hypermomentum1,hypermomentum2,hypermomentum3}. Equation~\eqref{203} imposes the symmetry of $\Theta^{\mu\nu}$ and hence the condition $\Theta_{[ab]}=0$. The general hypermomentum definition for any teleparallel gravity theory is exactly~\cite{hypermomentum1,hypermomentum2,hypermomentum3}:
\begin{align}\label{203c}
	\mathfrak{T}^{~\mu\nu}_{\sigma}= \frac{2}{h}\frac{\delta\,\left(h\,\mathcal{L}_{Matter}\right)}{\delta\,\Gamma^{\sigma}_{\mu\nu}} ,
\end{align}
where $\Gamma^{\sigma}_{\mu\nu}$ is the Weitzenbock connection. The hypermomentum $\mathfrak{T}^{~\mu\nu}_{\sigma}$ conservation law is expressed as ${\overset{\ \circ}{\nabla}}\,_{\nu}\left(h\,\mathfrak{T}^{~\mu\nu}_{\sigma}\right)=0$. In addition, there is a specific expression of Equation~\eqref{203c} for teleparallel $F(T)$-type gravity theory~\cite{nonvacSSpaper,nonvacKSpaper,scalarfieldKS,scalarfieldTRW}.

In the current situation, the perfect fluid energy-momentum tensor is~\cite{nonvacSSpaper,nonvacKSpaper,roberthudsonSSpaper,coley03}:
\begin{align}\label{204}
	\Theta_{(ab)}=T_{ab} = \left(P(\rho)+\rho\right)\,u_a\,u_b+g_{ab}\,P(\rho),
\end{align}
where $u_a=(-1,0,0,0)$, $P(\rho)=P(\rho(t))$ is the EoS, $\rho=\rho(t)$ is the time-dependent fluid density and $g_{ab}$ is the gauge metric. All of these previous quantities are time-dependent only in the cosmological geometry we are going to study.

For the scalar field $F(T,B)$ action integral, $\mathcal{L}_{matter}$ will be replaced by the scalar field {Lagrangian}~{\cite{Bahamonde:2021gfp,coleylandrygholami,leonftbcosmo1,scalar1,scalar2,scalarfieldTRW,cosmofluidsbohmer}}: 
\begin{align}\label{205}
	\mathcal{L}_{Scalar} = \frac{h}{2}{\overset{\ \circ}{\nabla}}\,_{\nu}\phi\,\overset{\ \circ}{\nabla}\,^{\nu}\phi -h\,V\left(\phi\right).
\end{align}

Applying Equation~\eqref{203} to Equation~\eqref{205} will lead to a {Klein--Gordon-like (KG-like) differential equation, and this result} will be shown in Section~\ref{sect6}~\cite{coleylandrygholami}. We will then find the $P_{\phi}$ and $\rho_{\phi}$ fluid equivalent for easily using the same form of FEs. We note that Equation~\eqref{205} allowing $P_{\phi}$ and $\rho_{\phi}$ expressions uses the covariant derivatives ${\overset{\ \circ}{\nabla}}_{\nu}$ because this last equation has to satisfy the GR conservation laws defined in Equation~\eqref{203}. By this approach, we will be able to compare the scalar field source teleparallel $F(T,B)$ solutions with those for perfect fluids, especially for the DE quintessence, the phantom energy, the bouncing and the quintom physical model purposes~\cite{cosmobaha,cosmoftbmodels,ftbcosmo2,ftbcosmo3,ftbcosmo4,ftbcosmo5,ftbcosmo6}.

\subsection{Teleparallel $F(T,B)$ FLRW Geometry}\label{sect23}

\noindent In this paper, we will use the same orthonormal coframe expression as~\cite{preprint,coleylandrygholami,scalarfieldTRW}:
\begin{align}
	h^a_{\;\;\mu} = Diag\left[1, a(t)\,\left(1-k\,r^2\right)^{-1/2},\,a(t)\,r,\, a(t)\,r\,\sin\theta\right].
\end{align}

The spin-connection components will also be~\cite{preprint,coleylandrygholami} :
\begin{align}
	\omega_{122} =& \omega_{133} = \omega_{144} =  W_1(t),   &\omega_{234} =& -\omega_{243} = \omega_{342} = W_2(t), 
	\nonumber \\
	\omega_{233} =& \omega_{244} = - \frac{\sqrt{1-kr^2}}{a(t)r},  &\omega_{344} =&  \frac{\cos(\theta)}{a(t) r \sin(\theta)}, \label{Con:FLRW} 
\end{align}
where $W_1$ and $W_2$ are functions depending on $k$-parameter for the inertial effects {and} are defined as {\cite{preprint,coleylandrygholami,scalarfieldTRW}}:

\begin{enumerate}
	\item $k=0$: $W_1=W_2=0$,
	\item $k=+1$: $W_1=0$ and $W_2(t)=\pm\,\frac{1}{a(t)}$,	
	\item $k={-1}$: $W_1(t)=\pm\,\frac{1}{a(t)}$ and $W_2=0$.		
\end{enumerate}

For any $W_1$ and $W_2$, we will obtain the same symmetric set of FEs to solve for each subcase depending on the $k$-parameter. The previous coframe and spin-connection expressions were found by solving Equations~\eqref{Intro:FS2} and imposing the null Riemann curvature condition (i.e., $R^a_{~b\mu\nu}=0$ as stated in ref.~\cite{Coley:2019zld}). These solutions were also used in the $F(T)$ TRW spacetime recent works~\cite{preprint,coleylandrygholami,scalarfieldTRW}.


\section{The General Perfect Fluid Field Equations}\label{sect3}

{We will define and solve the FEs for $F(T,B)$ theory, as defined by Equations~\eqref{202a} and~\eqref{202b}.} But we must define from Equation~\eqref{203} the energy-momentum conservation laws for a perfect fluid~\cite{coleylandrygholami,scalarfieldTRW}:

\vspace{-18pt}\begin{align}\label{301}
	\dot{\rho}+ 3\,H\,\left[\rho+P(\rho)\right]= 0 ,
\end{align}
{where $H=\frac{\dot{a}(t)}{a(t)}$ is the Hubble parameter and $P=P(\rho)$ is the EoS for {a} studied cosmological fluid.}

\subsection{$k=0$ Case}\label{sect31}

\subsubsection{General Equations}

For the $k=0$ case, the general FEs for any EoS are~\cite{Bahamonde:2021gfp}:
\vspace{-9pt}
\begin{align}
	\kappa\,\rho =& \,-3H\dot{F}_B + 3H^2\left(2F_T + 3F_B\right)+ 3\dot{H}F_B-\frac{F}{2},  \label{302a}
	\\
	\kappa(\rho+3P) =& \,F - 6H^2\left(2F_T + 3F_B\right)-6\dot{H}\left(F_T + F_B\right)- 3H\left(2\dot{F}_T + \dot{F}_B\right)+ 3\ddot{F}_B,  \label{302b}
\end{align}
where $F=F(T,B)$, $F_T=\frac{\partial F}{\partial T}$ and $F_B=\frac{\partial F}{\partial B}$. The torsion scalar $T(t)$, the curvature scalar $\overset{\ \circ}{R}(t)$ and the boundary variable $B(t)$ are, respectively, for $k=0$ (i.e., $B=\overset{\ \circ}{R}+T$):

\vspace{-18pt}
\begin{align}
	T(t) =& 6H^2 ,  \label{303a}
	\\
	\overset{\ \circ}{R}(t) =& 6\left(\dot{H}+ 2H^2\right), \label{303b}
	\\
	B(t) =& 6\left(\dot{H}+ 3H^2\right),  \label{303c}
\end{align}
where $\dot{H}=\frac{\ddot{a}}{a}-H^2$. 

\subsubsection{Power-Law Ansatz}

As in ref.~\cite{coleylandrygholami}, we will assume that $a(t)=a_0\,t^n$. Equations~\eqref{303a}--\eqref{303c} become:
\begin{align}
	T(t) =& \frac{6n^2}{t^2} ,  \label{403a}
	\\
	\overset{\ \circ}{R}(t) =& \frac{6n(2n-1)}{t^2}, \label{403b}
	\\
	B(t) =& \frac{6n(3n-1)}{t^2}.  \label{403c}
\end{align}

From Equations~\eqref{403a} and~\eqref{403c}, we can express the time coordinate $t$ in terms of $T$   {and/or $B$ as}:
\vspace{-6pt}\begin{align}\label{403d}
	t^2(T) = \frac{6n^2}{T} , \quad\text{and/or} \quad t^2(B) =\frac{6n(3n-1)}{B}.
\end{align}

Equations~\eqref{403d} constitute the simplest choice and the most frequently used in the {literature {\cite{cosmobaha,cosmoftbmodels,ftbcosmo2,ftbcosmo3,ftbcosmo4,ftbcosmo5,ftbcosmo6}}}. This last one has the main advantage of expressing separately the $F(T,B)$ functions under the $F(T,B)=F_1(T)+F_2(B)$ form. However, Equations~\eqref{403d} are not unique and there are a multitude of $t(T,B)$ relationships, which can be summarized under the form:
\begin{align}\label{403e}
	t(T,B)=&\,b_1\,\left(T+b_2\,B\right)^{-1/2}  ,
\end{align}
where $\left(b_1,\,b_2\right)$ are possible combinations. There are direct examples such as $\left(\sqrt{2n},\,-\frac{1}{3}\right)$ or $\left(3n\sqrt{2n},\,n\right)$ and several other ones. The $\left(b_1,\,b_2\right)$ possible sets are solutions of   {Equations~\eqref{403a} and~\eqref{403c}}. If $b_2=0$, we obtain in this case a pure $F(T)$ solution.

For all possible $t$ expressions, Equations~\eqref{302a}--\eqref{302b} become, for any fluid of   {EoS $P=P(\rho)$,}
\begin{align}
	\kappa\,\rho =& \,\frac{t^2\,\ddot{F}-9\left(n-\frac{2}{3}\right)\,t\,\dot{F}}{12\left(n-\frac{1}{3}\right)} -\frac{F}{2}  ,  \label{405a}
	\\
	\kappa\left(\rho+3P(\rho)\right) =& \,-\frac{\left[t^3\,\dddot{F}+(8-7n)\,t^2\,\ddot{F}+(10-9n-18n^2)\,t\,\dot{F}\right]}{12n\left(n-\frac{1}{3}\right)}+F  ,  \label{405b}
\end{align}
where $n\neq \left\lbrace 0,\,\frac{1}{3}\right\rbrace$, $F=F(t)$, $\dot{F}=F_t$ and $t$ is the time coordinate. By using the power-law ansatz $F(t)=F_0\,t^r$, Equations~\eqref{405a}--\eqref{405b} become algebraic relations:
\begin{align}
	\kappa_{eff}\,\rho =& \,\left[\frac{(r+5-9n)\,r}{12\left(n-\frac{1}{3}\right)} -\frac{1}{2}\right]\,t^r  ,  \label{405c}
	\\
	\kappa_{eff} \left(\rho+3P(\rho)\right) =& \,\left[1-\frac{\left[r^2+(5-7n)\,r+(4-2n-18n^2)\right]\,r}{12n\left(n-\frac{1}{3}\right)}\right]t^r  ,  \label{405d}
\end{align}
where $\kappa_{eff} =\frac{\kappa}{F_0}$. For most well-known EoS, we need to solve Equations~\eqref{405c}--\eqref{405d} for {finding} power-law $F(t)$ solutions and {then} $F(T,B)$ by substituting Equations~\eqref{403d} (or an Equation~\eqref{403e} relation) into $F(t)$. In addition, $n=0$ and $\frac{1}{3}$ singular cases in Equations~\eqref{405c}--\eqref{405d} are considered as special cases.


\subsection{$k=+1$ Case}\label{sect33}

\subsubsection{General Equations}

For the $k=+1$ case, the general FEs for any EoS $P(\rho)$ are~\cite{Bahamonde:2021gfp,preprint,coleylandrygholami}:
\begin{align}
	\kappa\,\rho =& \,-3H\dot{F}_B + 3H^2\left(2F_T + 3F_B\right)+ 3\dot{H}F_B-\frac{F}{2}, \label{361a}
	\\
	\kappa(\rho+3P) =& \,F - 6H^2\left(2F_T + 3F_B\right)-6\dot{H}\left(F_T + F_B\right)- 3H\left(2\dot{F}_T + \dot{F}_B\right)+ 3\ddot{F}_B + \frac{6}{a^2}F_T . \label{361b}
\end{align}

The torsion scalar $T(t)$, the curvature scalar $\overset{\ \circ}{R}(t)$ and the boundary variable $B(t)$ are, respectively, for $k=+1$:
\begin{align}
	T(t) =& 6\left(H^2-\frac{k}{a^2(t)}\right),  \label{362a}
	\\
	\overset{\ \circ}{R}(t) =& 6\left(\dot{H}+ 2H^2+ \frac{k}{a^2(t)}\right) , \label{362b}
	\\
	B(t) =& 6\left(\dot{H}+ 3H^2\right) . \label{362c}
\end{align}

\subsubsection{Power-Law Ansatz}

Assuming that $a(t)=a_0\,t^n$, Equations~\eqref{362a}--\eqref{362c} are:
\begin{align}
	T(t) =& 6\left(\frac{n^2}{t^2}-\frac{1}{a_0^2\,t^{2n}}\right),  \label{462a}
	\\
	\overset{\ \circ}{R}(t) =& 6\left(\frac{2(2n-1)}{t^2}+ \frac{1}{a_0^2\,t^{2n}}\right) , \label{462b}
	\\
	B(t) =& \frac{6n(3n-1)}{t^2} , \label{462c}
\end{align}

From Equations~\eqref{462a}--\eqref{462c}, the $t(T,B)$ expression for $n \neq \left\lbrace 0,\,\frac{1}{3}\right\rbrace$ is:
\begin{align}\label{462d}
	t(T,B) = \left(\frac{\delta_1}{a_0}\right)^{1/n}\left(\frac{n\,B}{(3n-1)}-\frac{T}{6}\right)^{-1/2n},
\end{align}
where $\delta_1=\pm 1$. However, we can express for $n=1$ that:
\begin{align}\label{462e}
	t^2(T)=\frac{6(a_0^2-1)}{a_0^2\,T} \quad\quad\text{and} \quad\quad t^2(B)=\frac{12}{B} ,
\end{align}
as for Equations~\eqref{403d}. The $t$-coordinate in terms of the $T$ and $B$ relationship is again not unique and a multitude of those are possible. Equations~\eqref{361a}--\eqref{361b} in terms of power-law ansatz are presented in Appendix~\ref{appen1b}.

\subsection{$k=-1$ case}\label{sect32}

\subsubsection{General Equations}

For the $k=-1$ case, the general FEs for any EoS $P(\rho)$ are {\cite{Bahamonde:2021gfp,preprint,coleylandrygholami}}:

	\begin{align}
		\kappa\,\rho =& \,-3\left(H+\frac{\delta}{a}\right)\dot{F}_B + 3H\left(H+\frac{\delta}{a}\right)\left(2F_T +3F_B\right)+ 3\dot{H}F_B-\frac{F}{2}, \label{331a}
		\\
		\kappa(\rho+3P) =& \,F - 6H\left(H+\frac{\delta}{a}\right)\left(2F_T + 3F_B\right)-6\dot{H}\left(F_T + F_B\right)-3H\left(H+\frac{\delta}{a}\right)\left(2\dot{F}_T +\dot{F}_B\right)
		\nonumber\\
		&\quad + 3\ddot{F}_B - \frac{6}{a^2}F_T. \label{331b}
	\end{align}

The torsion scalar $T(t)$, the curvature scalar $\overset{\ \circ}{R}(t)$ and the boundary variable $B(t)$ are, respectively, for $k=-1$:
\begin{align}
	T(t) =& 6\left(H+\frac{\delta\sqrt{-k}}{a(t)}\right)^2, \label{332a}
	\\		
	\overset{\ \circ}{R}(t) =&	6\left(\dot{H}+ 2H^2+ \frac{k}{a^2(t)}\right), \label{332b}
	\\
	B(t) =& 6\left(\dot{H}+ 3H^2+ 2H\frac{\delta\,\sqrt{-k}}{a(t)}\right) . \label{332c}
\end{align}

\subsubsection{Power-Law Ansatz}

Assuming that $a(t)=a_0\,t^n$, Equations~\eqref{332a}--\eqref{332c} are:
\begin{align}
	T(t) =& 6\left(\frac{n}{t}+\frac{\delta}{a_0\,t^n}\right)^2, \label{432a}
	\\		
	\overset{\ \circ}{R}(t) =&	6\left(\frac{n(2n-1)}{t^2}- \frac{1}{a_0^2\,t^{2n}}\right), \label{432b}
	\\
	B(t) =& 6\left(\frac{n(3n-1)}{t^2}+ \frac{2\delta n}{a_0\,t^{n+1}}\right), \label{432c}
\end{align}

From Equations~\eqref{432a} and~\eqref{432c}, the $t(T,B)$ expression for all values of $n$ is:
\begin{align}\label{432d}
	t(T,B) = \sqrt{6}n\frac{\sqrt{T}}{B}\left[1+\delta_1\sqrt{1+\left(1-\frac{1}{n}\right)\frac{B}{T}}\right],
\end{align}
where $\delta_1=\pm 1$. Once again, Equation~\eqref{432d} is not the only possible $t$-coordinate in terms of $T$ and $B$ relation. For $n=1$, we can separately express $t(T)$ and $t(B)$:
\begin{align}\label{432e}
	t^2(T)=\frac{6}{T}\left(1+\frac{\delta}{a_0}\right)^2 \quad\quad\text{and} \quad\quad t^2(B)=\frac{12}{B}\left(1+\frac{\delta}{a_0}\right) ,
\end{align}
as for Equations~\eqref{403d}. Equations~\eqref{331a}--\eqref{331b}, assuming a power-law ansatz, are presented in Appendix~\ref{appen1a}.

\subsection{Possible Solutions}

In Sections~\ref{sect4} and~\ref{sect5}, we will use the appropriate EoS $P(\rho)$ relations for each of the $k=0,\,-1,\,+1$ {cases} and by solving with the power-law ansatz approach. In Section~\ref{sect6}, we will use the pressure and density equivalent ${P_{\phi}}$ and ${\rho_{\phi}}$ {in terms of the scalar field }${\phi(t)}$ satisfying a cosmological KG equation. This is the most direct manner to set and solve the FEs for exact solutions. For all the coming sections, the first aim is to find possible $F(T,B)$ solutions in terms of Equations~\eqref{403d},~\eqref{403e},~\eqref{462d},~\eqref{462e},~\eqref{432d} and~\eqref{432e} (for $k=0,$ $-1$ and $+1$ cases, respectively). The possible new teleparallel $F(T,B)$ solutions can also be compared with those from the recent literature.


\section{Linear Perfect Fluid \boldmath{$F(T,B)$} Solutions}\label{sect4}

{For a linear perfect fluid of EoS $P(\rho)=\alpha\,\rho$ where $-1<\alpha \leq 1$, Equation~\eqref{301} becomes~{\cite{coleylandrygholami}}:}
\begin{align}\label{401}
	\dot{\rho}+ 3\,H\,\left(1+\alpha\right)\,\rho= 0.
\end{align}

By applying a power-law ansatz as $a(t)=a_0\,t^n$, we obtain as a solution:
\begin{align}\label{402}
	\dot{\rho}+ \frac{3n}{t}\,\left(1+\alpha\right)\,\rho= 0, \quad\quad \Rightarrow\,\rho(t)=\rho(0)\,t^{-3n\left(1+\alpha\right)},
\end{align}
where $H=\frac{n}{t}$, $\dot{H}=-\frac{n}{t^2}$. Equation~\eqref{402} shows a decreasing fluid density for positive values of $n$ (expanding universe).

\subsection{$k=0$ Case}\label{sect41}

\begin{enumerate}
	\item {\textbf{General solution:}}
	By putting Equations~\eqref{405c}--\eqref{405d} together for a power-law solution $F(t)=F_0\,t^r$, we will find as characteristic equation:
	\small
	\begin{align}
		0 =& \,r^3+\left(5+3n\left(\alpha-2\right)\right)\,r^2+\left(4+3n\left(1+5\alpha\right)-27n^2\left(1+\alpha\right)\right)r-18\left(1+\alpha\right)n\left(n-\frac{1}{3}\right)  .  \label{410}
	\end{align}
	\normalsize
	
	{The only real-valued ($\mathbb{R}$-valued) solution to Equation~\eqref{410} is (i.e., negative cubic~discriminant):}
	\small
	
		\begin{align}\label{410a}
			& r\left(\alpha,n\right)
			\nonumber\\
			=& \frac{1}{6}\Bigg[-1620 \alpha^{2} n^{3}+540 n^{2} \alpha^{2}+324 \alpha  n^{3}-1512 n^{2} \alpha +7560 n^{3}+684 n \alpha -7884 n^{2}
			+2628 n-280
			\nonumber\\
			&\; -216 \alpha^{3} n^{3}+36 \Bigg(-2187 n^{6} \alpha^{4}+1782 n^{5} \alpha^{4}-21870 n^{6} \alpha^{3}-459 n^{4} \alpha^{4}+24,138 n^{5} \alpha^{3}-72,171 n^{6} \alpha^{2}
			\nonumber\\
			&\;-9666 n^{4} \alpha^{3}+90,882 n^{5} \alpha^{2}-87,480 n^{6} \alpha +1350 \alpha^{3} n^{3}-43038 n^{4} \alpha^{2}+116,478 n^{5} \alpha -34,992 n^{6}
			\nonumber\\
			&\; +8586 \alpha^{2} n^{3}-60,210 n^{4} \alpha +47952 n^{5}-531 n^{2} \alpha^{2}+14,706 \alpha  n^{3}-30,267 n^{4}-1686 n^{2} \alpha +12,078 n^{3}
			\nonumber\\
			&\;+80 n \alpha -3363 n^{2}+592 n-48\Bigg)^{1/2}\Bigg]^{\frac{1}{3}}-6 \left(-5 n^{2} \alpha +\frac{5}{3} n \alpha -13 n^{2}+\frac{23}{3} n-\frac{13}{9}-n^{2} \alpha^{2}\right)
			\\
			&\;\times\,\Bigg[-1620 \alpha^{2} n^{3}+540 n^{2} \alpha^{2}+324 \alpha  n^{3} -1512 n^{2} \alpha +7560 n^{3}+684 n \alpha -7884 n^{2}+2628 n
			\nonumber\\
			&\;-280-216 \alpha^{3} n^{3}+36 \Bigg( -2187 n^{6} \alpha^{4}+1782 n^{5} \alpha^{4}-21,870 n^{6} \alpha^{3}-459 n^{4} \alpha^{4}+24,138 n^{5} \alpha^{3}
			\nonumber\\
			&\;-72171 n^{6} \alpha^{2}-9666 n^{4} \alpha^{3}+90,882 n^{5} \alpha^{2} -87,480 n^{6} \alpha +1350 \alpha^{3} n^{3}-43,038 n^{4} \alpha^{2}+116,478 n^{5} \alpha
			\nonumber\\
			&\; -34,992 n^{6}+8586 \alpha^{2} n^{3}-60,210 n^{4} \alpha  +47,952 n^{5}-531 n^{2} \alpha^{2}+14,706 \alpha  n^{3}-30267 n^{4}-1686 n^{2} \alpha 
			\nonumber\\
			&\;+12,078 n^{3}+80 n \alpha -3363 n^{2} +592 n-48\Bigg)^{1/2}\Bigg]^{-\frac{1}{3}} +n(2-\alpha) -\frac{5}{3} .\nonumber
		\end{align}
	\normalsize
	By using Equations~\eqref{403d}, the general teleparallel $F(T,B)$ solution is:
	\begin{align}\label{411}
		F(T,B) =&-\Lambda_0 \, {\delta_{n,\frac{1}{3}}\delta_{-1,\alpha}}+ c_1 \,T^{-\frac{r\left(\alpha,n\right)}{2}}+c_2\,B^{-\frac{r\left(\alpha,n\right)}{2}},
	\end{align}
	where $\Lambda_0$ is the cosmological constant, {$\delta_{n,\frac{1}{3}}$ and $\delta_{-1,\alpha}$ are the Kronecker delta terms,} $c_1$, $c_2$ are arbitrary constant and $r\left(\alpha,n\right)$ is Equation~\eqref{410a}. Equation~\eqref{411} is possible because the definitions of $t(T)$ and $t(B)$ are both equivalent, as seen in Equations~\eqref{403d}. The teleparallel $F(T,B)$ solution can be expressed in terms of Equation~\eqref{403e}:
	\begin{align}\label{411a}
		F(T,B) =&-\Lambda_0 \, {\delta_{n,\frac{1}{3}}\delta_{-1,\alpha}}+ c_1 \,\left(T+b_2\,B\right)^{-\frac{r\left(\alpha,n\right)}{2}},
	\end{align}
	where $b_2=-\frac{1}{3}$, $n$ or other possible values satisfying Equations~\eqref{403a} and~\eqref{403c}.

	Therefore, we can also find some specific teleparallel $F(T,B)$ solutions for some limit values by using Equations~\eqref{403d},~\eqref{410}--\eqref{411}:
	\begin{enumerate}
		\item {{${\bf n\rightarrow \frac{1}{3}}$:} Equation~\eqref{410} simplifies into $r^2+\left(3+\alpha\right)\,r+2\left(1+\alpha\right)=0$ and $r=0$, and then yields:}
		\small
		\begin{align}\label{412a}
			F(T,B)=F(T)=-\Lambda_0 +c_1\,T+c_2\,T^{\left(1+\alpha\right)/2}.
		\end{align}
		\normalsize
		
		\item {${\bf n\gg 1}$} (Large $n$): By applying $n\gg 1$ approximations and keeping only $n^2$ terms for very large $n$ limit (assuming $n \geq 20$ in this paper), Equation~\eqref{410} becomes $0\approx \,-27n^2\left(1+\alpha\right)\left(r+\frac{2}{3}\right)$ and the $F(T,B)$ solution is:
		\small
		\begin{align}\label{412d}
			F(T,B)\,\rightarrow\,-\Lambda_0 \,{\delta_{-1,\alpha}}+c_1\,T^{1/3}+c_2\,B^{1/3} .
		\end{align}
		\normalsize
	\end{enumerate}

	\item {$ \alpha \rightarrow {\bf -\frac{1}{3}}$ \textbf{(quintessence limit) solutions:}} By setting $\alpha \rightarrow -\frac{1}{3}$, Equation~\eqref{410} and the solution become:
	\small
	
		\begin{align}\label{410d}
			0=&  r^3 +(5-7n)r^2+(4-2n-18n^2) r -12n\left(n-\frac{1}{3}\right) ,
			\nonumber\\
			\Rightarrow\,r\left(-\frac{1}{3},n\right)=&\frac{(7n-5)}{3}+\frac{1}{3}\Bigg(910 n^{3}-915 n^{2}+300 n-35
			\nonumber\\
			&\,+3 \sqrt{-29,403 n^{6}+41,292 n^{5}-32,910 n^{4}+18,180 n^{3}-6435 n^{2}+1272 n-108}\Bigg)^{\frac{1}{3}}
			\\
			&\,+\frac{1}{3} \left(103n^{2}-64 n+13\right)\Bigg(910 n^{3}-915 n^{2}+300 n-35
			\nonumber\\
			&\, +3 \sqrt{-29,403 n^{6}+41,292 n^{5}-32,910 n^{4}+18,180 n^{3}-6435 n^{2}+1272 n-108}\Bigg)^{-\frac{1}{3}}.\nonumber
		\end{align}
	\normalsize
	
	In terms of Equations~\eqref{403d} and~\eqref{403e}, the $F(T,B)$ solutions will be expressed under Equations~\eqref{411} and~\eqref{411a}, respectively. But we find teleparallel $F(T,B)$ solutions with some specific values of $n$ by using the Equation~\eqref{410d} roots. In terms of Equations~\eqref{403d}, the Equation~\eqref{411} $F(T,B)$ solutions are:
	\begin{enumerate}
		\item {${\bf n=\frac{1}{6}}$:} 
		\begin{align}\label{414}
			F(T,B)=& c_{1}\,\sqrt{T} + c_2\,\sqrt{-B} + c_3\,T^{\frac{17+\sqrt{241}}{24}} + c_4\, (-B)^{\frac{17+\sqrt{241}}{24}}+ c_5\,T^{\frac{17-\sqrt{241}}{24}} 
			\nonumber\\
			&\,+ c_{6}\,(-B)^{\frac{17-\sqrt{241}}{24}} ,
		\end{align}
		where $B \leq 0$.
		
		\item {${\bf n\rightarrow \frac{1}{3}}$: }
		\begin{align}\label{415}
			F(T,B)\rightarrow F(T)=-\Lambda_0+ c_{1}\,T +c_2\,T^{1/3}  .
		\end{align}
		
		\item {${\bf n=\frac{1}{2}}$: }
		\begin{align}\label{416}
			F(T,B)=& c_1\,T+c_2\,B+c_3\,T^{-1/2}+c_4\,B^{-1/2}+ c_5\,T^{1/4}+c_6\,B^{1/4} .
		\end{align}
		
		\item {${\bf n\gg 1}$:} By taking the $n\gg 1$ approximation and keeping only $n^2$ terms, we find that the $F(T,B)$ solution is exactly Equation~\eqref{412d} for very large $n$ limit. Note that Equation~\eqref{412d} is an $\alpha$-independent $F(T,B)$ solution.
		
	\end{enumerate}
	
\end{enumerate}

 {For each previous new teleparallel $F(T,B)$ solution, we need to determine the $c_1$ to $c_6$ arbitrary constants depending on the other cosmological conditions. Some of the previous teleparallel $F(T,B)$ solutions are on the $F(T,B)=F_1(T)+F_2(B)$ form by using   {Equations~\eqref{403d}} $t$~versus $T$ and/or $B$ relationships as in several recent papers (see refs.~\cite{leonftbcosmo1,leonftbcosmo2,leon2,leon3} and references within). {A large number of the new $F(T,B)$ solutions described by Equations~\eqref{411}--\eqref{416}} are similar to recent solutions found in refs.~\cite{cosmobaha,cosmoftbmodels,cosmoftbenergy,ftbcosmo2,ftbcosmo3,ftbcosmo4,ftbcosmo5,ftbcosmo6} and provide generalizations of those recent literature solutions. Most of the previous new solutions will be useful to simulate the cosmological bounce, the DE quintessence and in some circumstances, the phantom energy physical processes. In addition, the new $F(T,B)$ solutions share common points with results from recent papers, but we can use some specific observational data such as redshift and Hubble parameter measurements and then apply the tensor-to-scalar ratio technique for comparing the new $F(T,B)$ solutions, fitting with the data in refs.~\cite{dixit,ftbcosmo3}.}  We can see that for $ n\rightarrow \frac{1}{3}$ with Equations~\eqref{403d}, the solution $F(T,B)=F_1(T)$ and $F_2(B)=0$ in this specific case. The other $F(T,B)$ solutions will be based on the Equation~\eqref{403e} form of the $t(T,B)$ relation and lead to another new class of teleparallel $F(T,B)$ solutions.

{On the parameter choices, the values $n=\frac{1}{2}$ and $\frac{1}{6}$ allow the highlighting of the differences between the $n<\frac{1}{3}$ and $n>\frac{1}{3}$ situations. The $n\gg 1$ ($n=20$ and beyond) cases show the very fast universe expansion scenario limit situations. The case $n=1$ constitutes the boundary between the slow and the fast universe expansion according to the literature and will be recurrent throughout the current paper's approach. Moreover, using $\alpha=-\frac{1}{3}$ constitutes the dark energy quintessence $\alpha$-parameter higher limit, and the previous results highlight this significance.} The previous $F(T,B)$ solutions are all new without exception {and beyond the current parameter choices}.

\subsubsection{Teleparallel Robertson--Walker Generalization}\label{TRWgeneralization}

In ref.~\cite{coleylandrygholami}, we also found for the same linear perfect EoS and $a(t)=a_0t^n$ ansatz that the general TRW $F(T)$ solution for the $k=0$ case is:
\begin{eqnarray}\label{TRW}
	\left(F(T)\right)_{TRW} &=& {\frac{C\,T^{\left[\frac{3(1+\alpha)n}{2}\right]}}{1-3(1+\alpha)n}} + b_1\,\sqrt{T} - \Lambda_0.
\end{eqnarray}

We can generalize Equation~\eqref{TRW} to $F(T,B)=\left(F(T)\right)_{TRW}+F_2(B)$. By substituting into Equations~\eqref{302a}--\eqref{302b} and putting them together, we find as unified FE:
\begin{align}
	0=	\frac{(1+\alpha)}{2}\left[\Lambda_0-F_2(B)+B\,F_2'(B)\right]+\frac{(n\alpha-1)}{n(3n-1)}B^2\,F_2''(B)-\frac{2}{3n(3n-1)}\,B^3\,F_2'''(B).  \label{TRWa}
\end{align}

Equation~\eqref{TRWa} is the differential equation to solve for any generalized TRW solution by finding the possible $F_2(B)$ solutions. By using a power-law solution ansatz as $F_2(B)=F_{20}\,B^q$, the homogeneous Equation~\eqref{TRWa} becomes an algebraic relation yielding:
\begin{align}
	0=&	(q-1)\left[q^2-\frac{(3n\alpha+1)}{2}\,q-\frac{3}{4}\,n(3n-1)(1+\alpha)\right],  
	\nonumber\\
	\Rightarrow\quad q_{\pm}(\alpha,n)=&\,\left(\frac{3n \alpha+1}{4}\right)\pm \sqrt{\left(\frac{3n \alpha+1}{4}\right)^2+\frac{3}{4}\,n(3n-1)(1+\alpha)} ,\label{TRWb}
\end{align}
and $q=1$. The TRW generalized $F(T,B)$ solution will be:
\begin{align}\label{TRWd}
	F(T,B)=& -\Lambda_0\,{\delta_{n,\frac{1}{3}}\delta_{-1,\alpha}+\frac{C\,T^{\left[\frac{3(1+\alpha)n}{2}\right]}}{1-3(1+\alpha)n}} + b_1\,\sqrt{T} + c_1\,B+ c_2\,B^{q_{+}(\alpha,n)}+c_3\,B^{q_{-}(\alpha,n)} ,
\end{align}
where $q_{\pm}(\alpha,n)$ is exactly Equation~\eqref{TRWb}, {$\delta_{n,\frac{1}{3}}$ and $\delta_{-1,\alpha}$ are the Kronecker delta terms. There are some common terms in Equation~\eqref{TRWd} with the $F(T,B)$ solutions found for power-law cosmology in ref.~\cite{cosmobaha}. However, there are double roots in the $B$-dependent terms of Equation~\eqref{TRWd} instead of a single root in ref.~\cite{cosmobaha}. In addition, there is no cosmological constant in the ref.~\cite{cosmobaha} solution, but this last term is necessary in $F(T,B)$ solutions for taking into account the $\alpha=-1$ DE cosmological perfect fluid for a realistic universe representation in our models. We can also compare Equation~\eqref{TRWd} by using observational data fitting techniques and we expect a better solution fitting in the current case~\cite{cosmobaha,dixit,ftbcosmo3}.} This TRW generalized solution is possible for a linear perfect fluid, but we will show in Section~\ref{sect51} that it is also possible to obtain such a generalization for a non-linear perfect fluid under specific conditions.

\subsection{$k=+1$ Case}\label{sect43}

The new solutions are found by substituting the $F(t)=F_0t^r$ ansatz into the FEs defined by Equations~\eqref{464a}--\eqref{464b} and then by putting those together. For most values of $n$, the solutions in Equations~\eqref{464a}--\eqref{464b} will only lead to cosmological constant $F(T,B)=-\Lambda_0$ with $\alpha=-1$ (and $r=0$ as the only root), a GR solution. But there are some specific situations which lead to new teleparallel $F(T,B)$ solutions by using the Equation~\eqref{462d} relationship for $t(T,B)$:
\begin{enumerate} 
	\item {${\bf n\rightarrow\,\frac{1}{3}}$:} By putting Equations~\eqref{464a}--\eqref{464b} together and setting $n=\frac{1}{3}+\epsilon$, we find by taking the $\epsilon\,\rightarrow\,0$ limit that the teleparallel $F(T,B)$ solution is:
	\begin{align}
		F(T,B) = -\Lambda_0+c_1\,\left(\frac{2B}{3\epsilon}-T\right)^{3}+c_2\,\left(\frac{2B}{3\epsilon}-T\right)^{3(1+\alpha)/2}. \label{481}
	\end{align}

	\item {${\bf n=1}$}: By putting Equations~\eqref{464a}--\eqref{464b} together and then setting $n=1$, we obtain:
		\begin{align}
			0=& \left(a_{0}^{2}-1\right) r^{3}+\left(\left(3 \alpha -1\right) a_{0}^{2}-3 \alpha -3\right) r^{2}+\left(8-4\left(3 \alpha +5\right) a_{0}^{2}\right) r-12\left(1+ \alpha\right) \left(a_{0}^{2}-1\right). \label{483}
		\end{align}
	\begin{enumerate} 
		\item  {{${\bf a_0=1}$}: The Equation~\eqref{483} simplifies as $0=r^{2}+3\left(\alpha+1\right) r$ and the $F(T,B)$ solution~is:}
		\begin{align}\label{483a}
			F(T,B) = -\Lambda_0+c_1\,\left(3B-T\right)^{\frac{3}{2}\left(\alpha+1\right)}.
		\end{align}
		
		\item  {{\textbf{General:}} The only $\mathbb{R}$-valued Equation~\eqref{483} solution is (i.e., negative cubic discriminant):}
		\small
		 {\begin{align}\label{483b}
				& r(\alpha,1) 
				\nonumber\\
				=& \frac{1}{3 \left(a_{0}^{2}-1\right)}\Bigg(-27 a_{0}^{6} \alpha^{3}-135 a_{0}^{6} \alpha^{2}-63 a_{0}^{6} \alpha +81 a_{0}^{4} \alpha^{3}+253 a_{0}^{6}+351 a_{0}^{4} \alpha^{2}+225 a_{0}^{4} \alpha
				\nonumber\\
				& -81 a_{0}^{2} \alpha^{3}-333 a_{0}^{4}-297 a_{0}^{2} \alpha^{2}+6 \Bigg(-486 a_{0}^{8} \alpha^{4}-3402 a_{0}^{8} \alpha^{3}-9153 a_{0}^{8} \alpha^{2}+1458 a_{0}^{6} \alpha^{4}
				\nonumber\\
				&-10,188 a_{0}^{8} \alpha +8910 a_{0}^{6} \alpha^{3}-4527 a_{0}^{8}+21546 a_{0}^{6} \alpha^{2}-1701 a_{0}^{4} \alpha^{4}+22,626 a_{0}^{6} \alpha -9234 a_{0}^{4} \alpha^{3}
				\nonumber\\
				&+10,452 a_{0}^{6} -20,682 a_{0}^{4} \alpha^{2}+972 a_{0}^{2} \alpha^{4}-20,898 a_{0}^{4} \alpha +4698 a_{0}^{2} \alpha^{3}-10,245 a_{0}^{4}+9666 a_{0}^{2} \alpha^{2}
				\nonumber\\
				&-243 \alpha^{4}+9270 a_{0}^{2} \alpha -972 \alpha^{3}+4866 a_{0}^{2}-1809 \alpha^{2}-1674 \alpha -978\Bigg)^{1/2}\, \left(a_{0}^{2}-1\right)
				\nonumber\\
				& 	-189 a_{0}^{2} \alpha +27 \alpha^{3}+171 a_{0}^{2}+81 \alpha^{2}+27 \alpha -27\Bigg)^{\frac{1}{3}}
				\\
				&+\frac{9 a_{0}^{4} \alpha^{2}+30 a_{0}^{4} \alpha +61 a_{0}^{4}-18 a_{0}^{2} \alpha^{2}-48 a_{0}^{2} \alpha -78 a_{0}^{2}+9 \alpha^{2}+18 \alpha +33}{3 \left(a_{0}^{2}-1\right)}
				\nonumber\\
				& \times\, \Bigg(-27 a_{0}^{6} \alpha^{3}-135 a_{0}^{6} \alpha^{2}-63 a_{0}^{6} \alpha +81 a_{0}^{4} \alpha^{3}+253 a_{0}^{6}+351 a_{0}^{4} \alpha^{2}+225 a_{0}^{4} \alpha
				\nonumber\\
				& -81 a_{0}^{2} \alpha^{3}-333 a_{0}^{4}-297 a_{0}^{2} \alpha^{2}+6 \Bigg(-486 a_{0}^{8} \alpha^{4}-3402 a_{0}^{8} \alpha^{3}-9153 a_{0}^{8} \alpha^{2}+1458 a_{0}^{6} \alpha^{4}
				\nonumber\\
				&-10,188 a_{0}^{8} \alpha +8910 a_{0}^{6} \alpha^{3}-4527 a_{0}^{8}+21546 a_{0}^{6} \alpha^{2}-1701 a_{0}^{4} \alpha^{4}+22,626 a_{0}^{6} \alpha -9234 a_{0}^{4} \alpha^{3}
				\nonumber\\
				&+10,452 a_{0}^{6}-20,682 a_{0}^{4} \alpha^{2}+972 a_{0}^{2} \alpha^{4}-20,898 a_{0}^{4} \alpha +4698 a_{0}^{2} \alpha^{3}-10245 a_{0}^{4}+9666 a_{0}^{2} \alpha^{2}
				\nonumber\\
				& -243 \alpha^{4}+9270 a_{0}^{2} \alpha -972 \alpha^{3}+4866 a_{0}^{2}-1809 \alpha^{2}-1674 \alpha -978\Bigg)^{1/2}\, \left(a_{0}^{2}-1\right)
				\nonumber\\
				&-189 a_{0}^{2} \alpha +27 \alpha^{3}+171 a_{0}^{2}+81 \alpha^{2}+27 \alpha -27\Bigg)^{-\frac{1}{3}}-\frac{3 a_{0}^{2} \alpha -a_{0}^{2}-3 \alpha -3}{3 \left(a_{0}^{2}-1\right)} .\nonumber
		\end{align}}
		\normalsize	
		
		\item  {$ \alpha={\bf -\frac{1}{3}}$:} Equation~\eqref{483b} will simplify as:
		\small
		
			\begin{align}\label{483c}
				r\left(-\frac{1}{3},1\right)=& \frac{1}{3 \left(a_{0}^{2}-1\right)}\Bigg(260 a_{0}^{6}-372 a_{0}^{4}+12 \sqrt{-3 \left(13 a_{0}^{4}-16 a_{0}^{2}+7\right)^{2}}\, \left(a_{0}^{2}-1\right)+204 a_{0}^{2}
				\nonumber\\
				&-28\Bigg)^{\frac{1}{3}}+\frac{4 \left(13 a_{0}^{4}-16 a_{0}^{2}+7\right)}{3 \left(a_{0}^{2}-1\right)} \Bigg(260 a_{0}^{6}-372 a_{0}^{4}
				+12 \sqrt{-3 \left(13 a_{0}^{4}-16 a_{0}^{2}+7\right)^{2}}\, 
				\nonumber\\
				&\times\,\left(a_{0}^{2}-1\right)+204 a_{0}^{2}-28\Bigg)^{-\frac{1}{3}}+\frac{2 \left(a_{0}^{2}+1\right)}{3 \left(a_{0}^{2}-1\right)} .
			\end{align}
		\normalsize
	\end{enumerate}	
	
	The final teleparallel $F(T,B)$ solution form for the two last subcases in terms of Equation~\eqref{462d} is:
	\begin{align}
		F(T,B)= -\Lambda_0 \, {\delta_{a_0^2,1}\delta_{-1,\alpha}}+c_1\,(3B-T)^{-r(\alpha,1)/2},
	\end{align}
	where {$\delta_{a_0^2,1}$ and $\delta_{-1,\alpha}$ are the Kronecker delta terms}, $r(\alpha,1)$ is Equation~\eqref{483b} for the general case and Equation~\eqref{483c} for the $\alpha= -\frac{1}{3}$ case.
	
	\item {${\bf n\gg 1}$}: By putting Equations~\eqref{464a}--\eqref{464b} together, applying the $F(t)=F_0\,t^r$ ansatz and taking the large $n$ limit, we find a simplified equation and the only $\mathbb{R}$-valued solution is:
	\small
	
	\vspace{-20pt} 
		\begin{align}\label{484}
		    & 0\approx \,	r^3+(60\alpha-115)r^2-\left(10,736+10,500\alpha\right)r -7080\left(1+\alpha\right)  , 
			\nonumber\\
			\Rightarrow\quad & r(\alpha,\infty)\approx  \frac{1}{3}\Bigg(-1,593,000 \alpha^{2}+250,110 \alpha +7,172,335-216,000 \alpha^{3}
			+18 \Bigg(-8,396,190,000 \alpha^{4}
			\nonumber\\
			&-82,748,385,000 \alpha^{3} -270,933,700,275 \alpha^{2}-327,219,189,050 \alpha -130,674,314,338\Bigg)^{1/2}\Bigg)^{\frac{1}{3}}
			\nonumber\\
			&  + \Bigg(5900 \alpha +\frac{45,433}{3}+1200 \alpha^{2}\Bigg)\Bigg(-1,593,000 \alpha^{2}+250,110 \alpha +7,172,335-216,000 \alpha^{3}
			\nonumber\\
			& +18 \Bigg(-8,396,190,000 \alpha^{4}-82,748,385,000 \alpha^{3}-270,933,700,275 \alpha^{2}-327,219,189,050 \alpha 
			\nonumber\\
			& -130,674,314,338\Bigg)^{1/2}\Bigg)^{-\frac{1}{3}} -20 \alpha +\frac{115}{3}. 
		\end{align}
	\normalsize
	
	The $F(T,B)$ solution is:
	\begin{align}
		F(T,B) \approx & \,-\Lambda_0 \,{\delta_{a_0^2,1}\delta_{-1,\alpha}}+c_1\,(2B-T)^{-r(\alpha,\infty)/40}, \label{485}
	\end{align}	
	where $r(\alpha,\infty)$ is the Equation~\eqref{484}.
\end{enumerate}

All the teleparallel $F(T,B)$ solutions for specific cases are new and yield to $(aB-T)^n$ terms (single or superposition). {Here again for $k=+1$ solutions, the cosmological constant term is ever necessary for $\alpha=-1$ DE cosmological perfect fluid representation in realistic universe models. As in the $k=0$ solutions, we can also compare the new $F(T,B)$ solutions of Equations~\eqref{481}--\eqref{485} with recent results (see refs.~\cite{cosmobaha,cosmoftbmodels,ftbcosmo2,ftbcosmo3,ftbcosmo4} to name a few) and with observational data techniques, as carried out in refs.~\cite{dixit,ftbcosmo3}.}

\subsection{$k=-1$ Case}\label{sect42}

By applying the $F(t)=F_0t^r$ ansatz to Equations~\eqref{434a}--\eqref{434b}, we find that most values of $n$ only yield to the cosmological constant solution $F(T,B)=-\Lambda_0$ with   {$\alpha=-1$}, a GR solution. As in Section~\ref{sect43}, this is because the FEs described by Equations~\eqref{434a}--\eqref{434b} are very huge and difficult to solve exactly with the several and different types of terms. By setting $a_0=1$ and some specific values of $n$ in   {Equations~\eqref{434a}--\eqref{434b}}, we find new teleparallel $F(T,B)$ solutions in terms of   {Equation~\eqref{432d}} for the cases:
\begin{enumerate} 
	\item {${\bf n=1}$}: We find the simplified $F(T,B)$ solution:
	\begin{align}\label{490}
		F(T,B)=& -\Lambda_0 \,{\delta_{-1,\alpha}} + c_1\,\frac{B^2}{T}.
	\end{align}

	\item  {${\bf n\gg 1}$}: The unified and approximated algebraic equation is:
	\small
	 {\begin{align}\label{491}
			& 0\approx \,  r^{3}+\left(60 \alpha +23\right) r^{2}-\left(-10,500 \alpha +10,460\right) r-7080 \left(\alpha+1\right),
			\nonumber\\
			\Rightarrow \quad & r(\alpha,\infty)= \frac{1}{3}\Bigg(-3,083,400 \alpha^{2}-3,910,590 \alpha -999,197-216,000 \alpha^{3}+18 \Bigg(-8,396,190,000 \alpha^{4}
			\nonumber\\
			&-121,229,271,000 \alpha^{3}-314,509,767,075 \alpha^{2}-298,870,619,780 \alpha -97,193,974,005\Bigg)^{1/2}\Bigg)^{\frac{1}{3}}
			\nonumber\\
			&-3 \left(-\frac{11,420}{3} \alpha -\frac{31,909}{9}-400 \alpha^{2}\right)\Bigg(-3,083,400 \alpha^{2}-3,910,590 \alpha -999,197-216,000 \alpha^{3}
			\\
			&+18 \Bigg(-8,396,190,000 \alpha^{4}-121,229,271,000 \alpha^{3}-314,509,767,075 \alpha^{2}-298,870,619,780 \alpha 
			\nonumber\\
			&-97,193,974,005\Bigg)^{1/2}\Bigg)^{-\frac{1}{3}}-20 \alpha -\frac{23}{3} ,\nonumber
	\end{align}} 
	\normalsize
	where we only keep the most dominating terms (the highest $n$-power terms). The new $F(T,B)$ solution is in terms of Equation~\eqref{432d} and for the $n\rightarrow \infty$ limit:
	\begin{align}\label{492}
		F(T,B)= \rightarrow\, -\Lambda_0 \,{\delta_{-1,\alpha}}+ c_1\,\left[\frac{\sqrt{T}+\delta_1\sqrt{T+B}}{B}\right]^{r(\alpha,\infty)} .
	\end{align}
\end{enumerate}

{Equation~\eqref{490} can be compared to a very similar solution found in ref.~\cite{ftbcosmo2}. At the $B\rightarrow T$ limit, Equation~\eqref{490} goes to a TEGR-like solution and Equation~\eqref{492} goes to a simple power-law expression in $T$ added by the cosmological constant. Equation~\eqref{490} might be a beginning point for a detailed cosmological data comparison-based study, feasible with some techniques, similar to those used in refs.~\cite{dixit,ftbcosmo3}.}


\section{Non-Linear Perfect Fluid \boldmath{$F(T,B)$} Solutions}\label{sect5}

For a non-linear perfect fluid of EoS $P(\rho)=\alpha\,\rho+\beta\,\rho^w$ where $\alpha\leq 1$ in this case (see refs.~\cite{nonvacSSpaper,nonvacKSpaper} {for good examples}), Equation~\eqref{301} becomes~\cite{coleylandrygholami,nonvacKSpaper}:
\begin{align}\label{501}
	\dot{\rho}+ 3\,H\,\left[(1+\alpha)\rho+\beta\,\rho^w\right]= 0.
\end{align}

By applying the ansatz $a(t)=a_0\,t^n$, we obtain as a solution:
\begin{align}\label{502}
	& \dot{\rho}+ \frac{3n}{t}\,\rho\,\left[(1+\alpha)+\beta\,\rho^{w-1}\right]= 0, 
	\nonumber\\
	& \Rightarrow\,\rho^{w-1}(t)=\frac{(1+\alpha)}{\left(\beta+(1+\alpha)\,\rho_0^{1-w}\right)\,t^{3n(1+\alpha)(w-1)}-\beta}.  
\end{align}

Because the FEs for each subcase in Section~\ref{sect3} will be non-linear, {the current study will be restricted to the $w=2$ case.} Equation~\eqref{502} for $w=2$ leads to:
\begin{align}\label{502a}
	\rho(t)=\frac{(1+\alpha)}{\left(\beta+(1+\alpha)\,\rho_0^{-1}\right)\,t^{3n(1+\alpha)}-\beta}
\end{align}

We will consider under such conditions a linear perfect fluid with a quadratic small correction $\beta \ll \alpha$  where the EoS is $P(\rho)=\alpha\,\rho+\beta\,\rho^2$ with $\alpha\leq 1$~\cite{nonvacSSpaper,nonvacKSpaper}.

\subsection{$k=0$ Case}\label{sect51}

For $k=0$ FEs and $F(t)=F_0\,t^r$ ansatz, Equation~\eqref{405c} remains identical. However, Equation~\eqref{405d}, by setting $w=2$ in the previous EoS and Equations~\eqref{501}--\eqref{502a}, becomes a quadratic FE in $\kappa\rho$:
\small
\begin{align}
	0=& \left(\kappa\rho\right)^2+\frac{\kappa\left(1+3\alpha\right)}{3\beta}\left(\kappa\rho\right) + \,\frac{\kappa}{3\beta}\left[\frac{\left[r^2+(5-7n)r+(4-2n-18n^2)\right]r}{12 n\left(n-\frac{1}{3}\right)}-1\right]\,c_0\,t^r  , 
	\nonumber\\
	\Rightarrow \, \kappa_{eff}\rho \approx & \,\frac{\kappa_{eff}}{6\beta}\left(1+3\alpha\right)\left(\delta_2-1\right)-\frac{\delta_2}{\left(1+3\alpha\right)}\left[\frac{\left[r^2+(5-7n)r+(4-2n-18n^2)\right]r}{12n\left(n-\frac{1}{3}\right)}-1\right]\,t^r,
	\label{503}
\end{align}
\normalsize
where $\kappa_{eff}\rho$ is also Equation~\eqref{405c}. For $\delta_2=+1$, we obtain exactly the linear perfect fluid FEs without any correction (see ref.~\cite{nonvacKSpaper}). For a real non-linear perfect fluid solutions, we set $\delta_2=-1$, and then by merging Equations~\eqref{405c} and~\eqref{503}, we find a linearized and approximated differential equation:
\small

\vspace{-12pt} 
	\begin{align}
		0 \approx & \left[r^3+\left(5-3n\left(\alpha+\frac{8}{3}\right)\right)\,r^2+\left(27n^2\,\left(\alpha-\frac{1}{3}\right)-15n\left(\alpha+\frac{7}{15}\right)+ 4\right)\,r+18n\left(n-\frac{1}{3}\right)\left(\alpha-\frac{1}{3}\right)\right]
		\nonumber\\
		&\;-\frac{4\kappa_{eff}}{\beta}n\left(n-\frac{1}{3}\right)\left(1+3\alpha\right)^2\,t^{-r}. \label{505} 
	\end{align}
\normalsize
The Equation~\eqref{505} leads to the following subcases:
\begin{enumerate} 
	\item {$\alpha \neq {\bf-\frac{1}{3}}$} and {${\bf n\,\rightarrow\,\frac{1}{3}}$}: Equation~\eqref{505} simplifies to $0 \approx \,\left(r+2\right) \left(r-\alpha +\frac{1}{3}\right) r$ and by using Equations~\eqref{403d}, the solution is:
	\begin{align}\label{505b}
		F(T,B)=F(T) \approx -\Lambda_0 + c_1\,T+c_2\,T^{\left(\frac{1}{6}-\frac{\alpha}{2}\right)}.
	\end{align}
	In terms of Equation~\eqref{403e}, the solution will be:
	\begin{align}\label{505c}
		F(T,B)=-\Lambda_0+	c_1\,\left(T+b_2\,B\right)+ c_2\left(T+b_2\,B\right)^{\left(\frac{1}{6}-\frac{\alpha}{2}\right)} ,
	\end{align}
	where $b_2$ satisfies Equations~\eqref{403a} and~\eqref{403c}.

	\item {$\alpha \rightarrow {\bf -\frac{1}{3}}$}: Equation~\eqref{505} leads exactly to  {Equations~\eqref{411} and~\eqref{410d} linear perfect fluid solutions and then to the $n$-valued subcase solutions described by}  Equations~\eqref{412d} and~\eqref{414}--\eqref{416}. The quintessence limit is the same for any perfect fluid with linear or non-linear EoS in this situation.
	
\end{enumerate}

As in Section~\ref{sect41}, the $n\,\rightarrow\,\frac{1}{3}$ case described by Equation~\eqref{505b} is going to a pure teleparallel $F(T)$ solution, also characterized by $F(T,B)=F_1(T)$ and $F_2(B)=0$. However, Equation~\eqref{505c} is a real teleparallel $F(T,B)$ solution, because we use a different $t(T,B)$ relation. In principle, the new previous $F(T,B)$ solutions can be easily compared to those from recent papers (see refs.~\cite{leonftbcosmo1,leonftbcosmo2,leon2,leon3} and references therein). As in Section~\ref{sect41}, the $n\,\rightarrow\,\frac{1}{3}$ subcase also leads to teleparallel $F(T)$ solutions as for the linear EoS situation when $\alpha \rightarrow -\frac{1}{3}$, because we have the same quintessence limit for any EoS with the linear dominating term. {The $F(T,B)$ solutions expressed by Equations~\eqref{505b} and~\eqref{505c} have some common characteristics with the Section~\ref{sect41} solutions, but they are specific to the current non-linear perfect fluid EoS. The non-linear perfect fluid EoS used in the current section is considered as an intermediate cosmological model in ref.~\cite{ftbcosmo3}. The authors studied in this paper the possible $F(T,B)$ solutions, but also the possible thermodynamic parameters, the redshift and the Hubble parameter values for observational comparison by tensor-to-scalar methods, as also done in a similar manner in ref.~\cite{dixit}. Therefore,} the previous $F(T,B)$ solutions are all new   {without exception}.

\subsubsection{{Teleparallel} Robertson--Walker Generalization}

 {As in Section~\ref{TRWgeneralization} and ref.~\cite{coleylandrygholami}, we will use Equation~\eqref{TRW} and generalize the {$F(T,B)=\left(F(T)\right)_{TRW}+F_2(B)$} approach for a non-linear EoS. By substituting into Equations~\eqref{302a}--\eqref{302b} and {then} by putting them together, we find the unified FE:}
\begin{align}\label{nonlineTRWDE}
	0 \approx &\frac{\kappa}{3\beta}\left(3\alpha+1\right)^2-C\left(3\alpha+1\right)\,T^{3n\left(1+\alpha\right)/2}  -\left(\frac{3\alpha-1}{2}\right)\left[F_2-B\,F_{2}'-\Lambda_0\right]
	\nonumber\\
	& \quad -\sqrt{\frac{6n}{3n-1}}\left(1+\frac{3\alpha}{2}\right)\,B^{1/2}\dot{F}_2'+3\ddot{F}_2',
\end{align}
where $F_{2}'=\frac{dF_2}{dB}$. Because we are looking for a pure $F_2(B)$, there are two possible solutions to achieve this {aim} with Equation~\eqref{nonlineTRWDE}:
\begin{itemize}
	\item {${\bf C=0}$}: Equation~\eqref{TRW} becomes $\left(F(T)\right)_{TRW} = b_1\,\sqrt{T} - \Lambda_0$ and then by applying the $F_2(B)=c_0\,B^q$ ansatz to Equation~\eqref{nonlineTRWDE}, we {find} as algebraic equation and solution:	
	\vspace{-12pt} 
		\begin{align}\label{nonlineTRWalgC0}
			& \frac{\kappa}{3\beta}\left(3\alpha+1\right)^2+\left(\frac{3\alpha-1}{2}\right)\Lambda_0 \approx (q-1)\,c_0\,B^q \left[\left(\frac{1-3\alpha}{2}\right)-\frac{n(2+3\alpha)q+(2q-1)q}{n(3n-1)}\right] ,
			\nonumber\\
			&\Rightarrow\quad	q_{\pm}(\alpha,n) = \left(\frac{1-n(2+3\alpha)}{4}\right)\left[1\pm \sqrt{1-\frac{4n(3n-1)(3\alpha-1)}{\left(1-n(2+3\alpha)\right)^2}}\right] ,
		\end{align}
	and $q=1$. The general $F(T,B)$ solution from Equation~\eqref{nonlineTRWalgC0} is:
	\begin{align}\label{C0solution}
		F(T,B)\approx &  -\Lambda_0 \,{\delta_{\frac{1}{3},\alpha}} +\frac{2\kappa\,(3\alpha+1)^2}{3\beta\,(3\alpha-1)}+ b_1\,\sqrt{T}  +c_1\,B +c_2\,B^{q_{+}(\alpha,n)} +c_3\,B^{q_{-}(\alpha,n)},
	\end{align}
	where $q_{\pm}(\alpha,n)$ is Equation~\eqref{nonlineTRWalgC0} {and ${\alpha \neq \frac{1}{3}}$.}
	
	\item {$\alpha={\bf -\frac{1}{3}}$}  {(quintessence limit): Equation~\eqref{TRW} becomes $\left(F(T)\right)_{TRW} =\frac{C\,T^n}{(1-2n)} + b_1\,\sqrt{T} - \Lambda_0$} where $n \neq \frac{1}{2}$ and by applying the same $F_2(B)=c_0\,B^q$ ansatz to the Equation~\eqref{nonlineTRWDE}, we find that:
	\begin{align}\label{nonlineTRWquintalg}
		\Lambda_0 \approx &  \frac{c_0\,B^q}{n(3n-1)}\,(q-1)\left[2q^2+(n-1)q-n(3n-1)\right],
	\end{align}
	where the roots are $q=1$, $n$ and $\frac{1-3n}{2}$. The general $F(T,B)$ solution is:
	\begin{align}\label{quintsol}
		F(T,B)\approx & -\Lambda_0 \,{\delta_{\frac{1}{3},n}} +\frac{C}{(1-2n)}\,T^n + b_1\,\sqrt{T} +c_1\,B+c_2\,B^n + c_3\,B^{\frac{1-3n}{2}}.
	\end{align}
	
\end{itemize}

In this subsection, we find the non-linear EoS possible TRW generalization from the ref.~\cite{coleylandrygholami} solution and we still obtain that $F(T,B)=F(T)+F_2(B)$ forms in each situation. {We can also compare Equations~\eqref{C0solution} and~\eqref{quintsol} to some solutions found in refs.~\cite{cosmobaha,cosmoftbmodels,ftbcosmo3} without any loss of generality and with observational data fitting techniques in the same manner as refs.~\cite{dixit,ftbcosmo3}. For this method, the observational data fitting techniques will allow for the non-linear fluid case to know the range of $\alpha$ values, but also to determine the quadratic term correction range of parameter values.} Indeed, Equations~\eqref{C0solution} and~\eqref{quintsol} are {really} new teleparallel $F(T,B)$ solutions.

\subsection{$k=+1$ and $-1$ Cases}\label{sect53}

\noindent {${\bf k=+1}$ \textbf{case:}} By using and merging Equations~\eqref{464a}--\eqref{464b} for the non-linear EoS and by using the $F(t)=F_0\,t^r$ ansatz, we find again that the most of $n$-values lead to cosmological solution $F(T,B)=-\Lambda_0$ under the condition $F_{0} =-\frac{2 \kappa}{3\beta}$. As in   {Sections~\ref{sect43} and~\ref{sect42}}, there are the teleparallel $F(T,B)$ solutions of Equations~\eqref{464a}--\eqref{464b} in terms of Equation~\eqref{462d} for the cases:
\begin{enumerate} 
	\item {${\bf n\rightarrow\,\frac{1}{3}}$}: The unified and simplified equation to solve is $0=r\left(r-\alpha +\frac{1}{3}\right) \left(r+2\right)$ and the $F(T,B)$ solution is:
	\begin{align}\label{580}
		F(T,B) = -\Lambda_0 + c_1\,\left(\frac{2B}{3\epsilon}-T\right)^{3}+c_2\,\left(\frac{2B}{3\epsilon}-T\right)^{(1-3\alpha)/2}.
	\end{align}
	
	\item {${\bf n=1}$}: There are teleparallel $F(T,B)$ solutions for Equations~\eqref{464a}--\eqref{464b} only in the limits {$\alpha \rightarrow\,{\bf -\frac{1}{3}}$} and/or $a_0\rightarrow\,\pm 1$:
	\begin{enumerate} 
		\item {$\alpha \rightarrow\,{\bf-\frac{1}{3}}$}: Equations~\eqref{464a}--\eqref{464b} simplify as:
		\small
		\vspace{-12pt} 
			\begin{align}\label{581}
				0 \approx & \left(a_0^2-1\right)r^{3}-2\left(a_0^2+1\right) r^{2}-8\left(2a_0^2-1\right) r-8\left(a_0^2-1\right) ,
				\nonumber\\
				\Rightarrow\quad r(a_0) =& \frac{1}{3 \left(a_{0}^{2}-1\right)}\Bigg(260 a_{0}^{6}-372 a_{0}^{4}+12 \sqrt{-3 \left(13 a_{0}^{4}-16 a_{0}^{2}+7\right)^{2}}\, \left(a_{0}^{2}-1\right)+204 a_{0}^{2}
				\nonumber\\
				&-28\Bigg)^{\frac{1}{3}} +\frac{4 \left(13 a_{0}^{4}-16 a_{0}^{2}+7\right)}{3 \left(a_{0}^{2}-1\right)} \Bigg(260 a_{0}^{6}-372 a_{0}^{4}
				+12 \sqrt{-3 \left(13 a_{0}^{4}-16 a_{0}^{2}+7\right)^{2}}\, 
				\nonumber\\
				&\times\,\left(a_{0}^{2}-1\right)+204 a_{0}^{2}-28\Bigg)^{-\frac{1}{3}}+\frac{2 \left(a_{0}^{2}+1\right)}{3 \left(a_{0}^{2}-1\right)}.
			\end{align}
		\normalsize
		The Equation~\eqref{581} solution yields:
		\begin{align}\label{582}
			F(T,B) =& -\Lambda_0\, {\delta_{a_0^2,1}}+ c_1\,\left(3B-T\right)^{-r(a_0)/2}.
		\end{align}
		
		\item {${\bf a_0=\pm 1}$} and $\alpha \neq {\bf-\frac{1}{3}}$: Equations~\eqref{464a}--\eqref{464b} simplify as $0=r(r+1-3\alpha)$ and yield:
		\begin{align}\label{583}
			F(T,B) =& -\Lambda_0+c_1\,\left(3B-T\right)^{(1-3\alpha)/2},
		\end{align}
		where $\alpha\neq \frac{1}{3}$.
		
		\item {${\bf a_0=\pm 1}$} and $ \alpha ={\bf -\frac{1}{3}}$: Equations~\eqref{464a}--\eqref{464b} simplify as $0=r(r+2)$ and yield:
		\begin{align}\label{584}
			F(T,B) =& -\Lambda_0+c_1\,\left(3B-T\right).
		\end{align}
		Equation~\eqref{584} is an Einstein--Cartan $F(T,B)$ solution in this case. By taking the $B\rightarrow\,T$ limit, we will recover the TEGR-like solution. 
	\end{enumerate}

\end{enumerate}

 \noindent {As in Section~\ref{sect43}}, the new previous $F(T,B)$ solutions are under the {$F(T,B) \sim c_1(a\,B-T)^{b}$} form or superposition of this form's terms. This type of term is a kind of power-law generalization compared to the $k=0$ cases where we obtain essentially some simple power-law combinations. {The previous new $F(T,B)$ solutions go further compared to results found in refs.~\cite{cosmobaha,cosmoftbmodels,cosmoftbenergy,ftbcosmo2,ftbcosmo3,ftbcosmo4,ftbcosmo5,ftbcosmo6} where there is no real equivalent of Equation~\eqref{580}--\eqref{584}. There is the opportunity to compare those solutions by using observational data techniques for parameter determination as used in refs.~\cite{dixit,ftbcosmo3}.}

{${\bf k=-1}$ \textbf{case:}} By merging and unifying Equations~\eqref{434a}--\eqref{434b}, and then performing the $F(t)=F_0t^r$ ansatz as in Section~\ref{sect53}, we find that the solutions are possible only by setting $a_0=1$. Most of the $n$-valued cases lead to the cosmological constant $F(T,B)=-\Lambda_0$ solution under the $F_{0} =-\frac{2 \kappa}{3\beta}$ condition as in Sections~\ref{sect43} {and}~\ref{sect42}. Therefore, we find a teleparallel $F(T,B)$ solution only for the {${\bf n=1}$} case. Equations~\eqref{434a}--\eqref{434b} lead to an unified FE where a teleparallel $F(T,B)$ solution is possible only by setting $\delta=-1$ and/or $\alpha=-\frac{1}{3}$. From this last consideration, we find that the $F(T,B)$ solution in terms of Equation~\eqref{432d} is exactly Equation~\eqref{490} as for the linear perfect fluid {equivalent solution}.


\section{Scalar Field \boldmath{$F(T,B)$} Solutions}\label{sect6}

\noindent The conservation law for scalar {field} EoS is~\cite{coleylandrygholami,leonftbcosmo1,leonftbcosmo2,scalarfieldTRW}:
\begin{align}\label{kleingordon}
	0= \ddot{\phi}+V'(\phi)+3\,H\,\dot{\phi}, 
\end{align}
where $V(\phi)$ is the scalar potential depending on the field ${\phi(t)}$. The $P_{\phi}$ and $\rho_{\phi}$ expressions in terms of $\phi(t)$ will be~\cite{coleylandrygholami,leonftbcosmo1,leonftbcosmo2,leon2,leon3,scalarfieldTRW,cosmofluidsbohmer,steinhardt1,steinhardt2}:
\begin{eqnarray}
	\rho_{\phi} = \frac{\dot{\phi}^2}{2}+V(\phi) \quad\quad \text{and}\quad\quad	P_{\phi}= \frac{\dot{\phi}^2}{2}-V(\phi) . \label{densitypressurephi}
\end{eqnarray}

In terms of the linear perfect fluid EoS ${P_{\phi}=\alpha_Q\,\rho_{\phi}}$, {the ratio $\alpha_Q$ is~\cite{steinhardt1,steinhardt2}}:
\begin{align}\label{Quintessenceindex}
	\alpha_Q =& \frac{P_{\phi}}{\rho_{\phi}} = \frac{\dot{\phi}^2-2V\left(\phi\right)}{\dot{\phi}^2+2V\left(\phi\right)},
\end{align}
where {the values define the possible DE physical processes:
	\begin{itemize}
		\item Quintessence $-1<\alpha_Q<-\frac{1}{3}$: This DE form explains the accelerating universe expansion by a fundamental scalar field~\cite{cosmoftbmodels,steinhardt1,steinhardt2,scalarfieldKS}.
		
		\item  {Cosmological constant $\alpha_Q=-1$: This case is often required for the DE impact representation in cosmological models. A constant $V\left(\phi\right)$ leads by default to this DE case.} 
		
		\item Phantom energy $\alpha_Q<-1$: This case can explain an uncontrolled accelerating universe expansion leading to the Big Rip physical process after some time delay~\cite{cosmoftbmodels,caldwell1,farnes,baumframpton,scalarfieldKS}. This type of model can also be explained by a scalar field.
		
		\item Quintom models: This case is a combination of Quintessence and Phantom energy models~\cite{quintom1,quintom2,quintom3,quintom4,quintomcoleytot,quintomteleparallel1}. Usually, this type of model is described by a two-scalar field action integral system. 
	\end{itemize}
	
	We should note that a comparative study was conducted between the parameters $\alpha_Q$ and the potential $V(\phi)$ by using the tracker solutions in ref.~\cite{steinhardt2}. This last method was used with a density $Q$ instead of $\phi$ inside a KG-like equation, like Equation~\eqref{kleingordon}. This was to determine the most relevant expressions of $V(\phi)$ satisfying Equations~\eqref{kleingordon}--\eqref{Quintessenceindex} and also for future observational data fitting purposes. 
	
}

Equations~\eqref{densitypressurephi} satisfy the Equation~\eqref{kleingordon} conservation law. From these equations, we are able to find the unified FEs for the $k=0,\,-1,\,+1$ {cases}. By first  assuming a power-law scalar field as $\phi(t)=p_0\,t^p$ where $p\,\in\,\mathbb{R}$ and still using the power-law ansatz for $a(t)$ as previously, we will find some possible $F(T,B)$ scalar field EoS equivalent solutions. We will then use an exponential $\phi(t)=p_0\,\exp\left(p\,t\right)$ ansatz (an infinite sum of power-laws terms) as a type of limit. The current section first aims to find possible $F(T,B)$ solutions for typical scalar field sources. We assume that the used scalar fields are $\mathbb{R}$-valued functions. For conservation laws {defined} by Equation~\eqref{kleingordon}, we find the following relations to satisfy:
\begin{enumerate} 
	\item {\textbf{Power-law scalar field:}} By substituting $\phi(t)=\phi_0\,t^p$ into Equation~\eqref{kleingordon}, we find that:
	\begin{align}\label{PLsol}
		0=& p_0^2\,p^2\,(p-1+3n)\,t^{2p-3} +\dot{V}(t),
		\nonumber\\	
		\Rightarrow V(\phi) =& V_0+\frac{(1-3n-p)\,p^2\,p_0^{2/p}}{2(p-1)}\,\phi^{2-2/p} ,
	\end{align}
	and then Equation~\eqref{Quintessenceindex} is:
	\begin{align}\label{quintPL}
		\alpha_Q =& -1+\frac{p^2 p_0^{2/p} \phi^{2-2/p}}{V_0-\frac{3n\,p^2\,p_0^{2/p}}{2(p-1)}\,\phi^{2-2/p} }.
	\end{align}
	
	For $p\gg 1$, Equation~\eqref{PLsol} will be as $V(\phi) \approx V_0-\frac{p^2}{2}\,\phi^{2}$ and then Equation~\eqref{quintPL} becomes $\alpha_Q \approx -1+\frac{p^2}{V_0}\phi^{2}$. {This large $p$ limit case when $V_0>0$ represents the ideal DE quintessence process scenario.}

	\item  {{\textbf{Exponential scalar field:}} By substituting $\phi(t)=p_0\,\exp\left(p\,t\right)$ into Equation~\eqref{kleingordon}, we find that:}
	\begin{align}\label{expsol}
		0=& p_0^2\,p^2\,(p-1+3n)\,t^{2p-3} +\dot{V}(t) ,
		\nonumber\\	
		\Rightarrow V(\phi) =& V_0-\frac{p^2}{2}\,\phi^2+3np^2\,p_0^2\,Ei\left(-2\ln\left(\frac{\phi}{p_0}\right)\right) ,
	\end{align}
	where $Ei\left(-2\ln\left(\frac{\phi}{p_0}\right)\right)$ is the exponential integral special function. The first   {two terms} of the Equation~\eqref{expsol} solution are the same as Equation~\eqref{PLsol} for the large $p$ limit. The last   {Equation~\eqref{expsol}} solution term is for the infinite power-law series bounding leading to the exponential scalar field. Equation~\eqref{Quintessenceindex} will be:
	\begin{align}\label{quintEXP}
		\alpha_Q =& -1+ \frac{p^2\,\phi^2}{V_0+3np^2\,p_0^2\,Ei\left(-2\ln\left(\frac{\phi}{p_0}\right)\right)}.
	\end{align}
	{This last case can also represent a quintessence scenario for a positive denominator value situation.

		\item {\textbf{Constant scalar field:}} By substituting $\phi=p_0$ ($p=0$ case) in Equations~\eqref{kleingordon}--\eqref{Quintessenceindex}, we find that $V(\phi)=V_0$, $P_{\phi}=-\rho_{\phi}=-V_0$ leading at the end to $\alpha_Q=-1$, the cosmological constant, a GR solution. This case also represents the limit between quintessence and phantom energy scenarios.}
	
\end{enumerate}

\subsection{$k=0$ Case}\label{sect61}

The terms of Equations~\eqref{405a}--\eqref{405b} with Equations~\eqref{densitypressurephi} are:
\begin{align}
	\kappa\,\left(\frac{\dot{\phi}^2}{2}+V(\phi)\right) =& \,\frac{t^2\,\ddot{F}-9\left(n-\frac{2}{3}\right)\,t\,\dot{F}}{12\left(n-\frac{1}{3}\right)} -\frac{F}{2} ,  \label{604a}
	\\
	\kappa\left(\dot{\phi}^2-V(\phi)\right) =& \,-\frac{\left[t^3\,\dddot{F}+(8-7n)\,t^2\,\ddot{F}+(10-9n-18n^2)\,t\,\dot{F}\right]}{24n\left(n-\frac{1}{3}\right)}  +\frac{F}{2} ,  \label{604b}
\end{align}
where $n\neq \left\lbrace 0,\,\frac{1}{3}\right\rbrace$. By merging Equations~\eqref{604a}--\eqref{604b}, and then by isolating the $\dot{\phi}^2$ term to eliminate $V(\phi)$, we find the unified FE for any scalar field $\phi(t)$:
\begin{align}\label{605}
	\kappa\,\dot{\phi}^2=& \,\frac{\left[-t^3\,\dddot{F}+(9n-8)\,t^2\,\ddot{F}+(21n-10)\,t\,\dot{F}\right]}{36n\left(n-\frac{1}{3}\right)} . 
\end{align}

\subsubsection{Power-Law Scalar Field Solutions}

By setting {a power-law scalar field} $\phi(t)=p_0\,t^p$ and a power-law ansatz $F(t)=F_0\,t^r$, Equation~\eqref{605} becomes:
\begin{align}\label{606}
	\kappa_{eff}\,p_0^2\,p^2\,t^{2p-2}=& \,\frac{\left[-r^2+(9n-5)\,r+12\left(n-\frac{1}{3}\right) \right]\,r\,t^r}{36n\left(n-\frac{1}{3}\right)} .  
\end{align}

For the $p_0=0$ case {(homogeneous-based)}, the Equation~\eqref{606} roots are $r=\frac{9 n-5}{2}\pm \frac{\sqrt{81 n^{2}-42 n+9}}{2}$ and $0$. For the $p_0 \neq 0$ case (source-based), the only possible solution is obtained by setting $p=\frac{r}{2}+1$ and then Equation~\eqref{606} becomes:
\small

	\begin{align}\label{608}
		0=&-r^{3}+\left(-9 \kappa  n^{2} p_{0}^{2}+3 \kappa  n p_{0}^{2}+9 n-5\right) r^{2}-36 \left(\kappa  n p_{0}^{2}-\frac{1}{3}\right) \left(n-\frac{1}{3}\right) r-36 \kappa  p_{0}^{2} n \left(n-\frac{1}{3}\right),
		\nonumber\\
		\Rightarrow\,r_p(n)=& \frac{1}{3}\,\Bigg(243 n+81 \kappa^{2} n^{3} p_{0}^{4}-1215 \kappa^{2} n^{4} p_{0}^{4}+27 \kappa^{2} n^{2} p_{0}^{4}-729 n^{2}-351 \kappa  n^{2} p_{0}^{2}+63 \kappa  n p_{0}^{2}+729 n^{3}
		\nonumber\\
		&-35-729 \kappa^{3} n^{6} p_{0}^{6}+729 \kappa^{3} n^{5} p_{0}^{6}-243 \kappa^{3} n^{4} p_{0}^{6}+27 \kappa^{3} n^{3} p_{0}^{6}+2187 \kappa^{2} n^{5} p_{0}^{4}-2187 \kappa  n^{4} p_{0}^{2}
		\nonumber\\
		&+1215 \kappa  n^{3} p_{0}^{2}+18 \Bigg(-729 \kappa^{3} n^{7} p_{0}^{6}+972 \kappa^{3} n^{6} p_{0}^{6}-486 \kappa^{3} n^{5} p_{0}^{6}+108 \kappa^{3} n^{4} p_{0}^{6}-9 \kappa^{3} n^{3} p_{0}^{6}
		\nonumber\\
		&+1215 \kappa^{2} n^{6} p_{0}^{4}-972 \kappa^{2} n^{5} p_{0}^{4}+135 \kappa^{2} n^{4} p_{0}^{4}+54 \kappa^{2} n^{3} p_{0}^{4}-12 \kappa^{2} n^{2} p_{0}^{4}-243 \kappa  n^{5} p_{0}^{2}+864 \kappa  n^{4} p_{0}^{2}
		\nonumber\\
		&-630 \kappa  n^{3} p_{0}^{2}+192 \kappa  n^{2} p_{0}^{2}-23 \kappa  n p_{0}^{2}-243 n^{4}+288 n^{3}-138 n^{2}+32 n-3\Bigg)^{1/2}\Bigg)^{\frac{1}{3}}
		\nonumber\\
		&-3 \left(-4 \kappa  n^{2} p_{0}^{2}-\frac{2}{3} \kappa  n p_{0}^{2}+6 n-\frac{13}{9}-9 \kappa^{2} n^{4} p_{0}^{4}+6 \kappa^{2} n^{3} p_{0}^{4}-\kappa^{2} n^{2} p_{0}^{4}+18 \kappa  n^{3} p_{0}^{2}-9 n^{2}\right)
		\\
		&\times\,\Bigg(243 n+81 \kappa^{2} n^{3} p_{0}^{4}-1215 \kappa^{2} n^{4} p_{0}^{4}+27 \kappa^{2} n^{2} p_{0}^{4}-729 n^{2}-351 \kappa  n^{2} p_{0}^{2}+63 \kappa  n p_{0}^{2}+729 n^{3}
		\nonumber\\
		&-35-729 \kappa^{3} n^{6} p_{0}^{6}+729 \kappa^{3} n^{5} p_{0}^{6}-243 \kappa^{3} n^{4} p_{0}^{6}+27 \kappa^{3} n^{3} p_{0}^{6}+2187 \kappa^{2} n^{5} p_{0}^{4}-2187 \kappa  n^{4} p_{0}^{2}
		\nonumber\\
		&+1215 \kappa  n^{3} p_{0}^{2}+18 \Bigg(-729 \kappa^{3} n^{7} p_{0}^{6}+972 \kappa^{3} n^{6} p_{0}^{6}-486 \kappa^{3} n^{5} p_{0}^{6}+108 \kappa^{3} n^{4} p_{0}^{6}-9 \kappa^{3} n^{3} p_{0}^{6}
		\nonumber\\
		&+1215 \kappa^{2} n^{6} p_{0}^{4}-972 \kappa^{2} n^{5} p_{0}^{4}+135 \kappa^{2} n^{4} p_{0}^{4}+54 \kappa^{2} n^{3} p_{0}^{4}-12 \kappa^{2} n^{2} p_{0}^{4}-243 \kappa  n^{5} p_{0}^{2}+864 \kappa  n^{4} p_{0}^{2}
		\nonumber\\
		&-630 \kappa  n^{3} p_{0}^{2}+192 \kappa  n^{2} p_{0}^{2}-23 \kappa  n p_{0}^{2}-243 n^{4}+288 n^{3}-138 n^{2}+32 n-3\Bigg)^{1/2}\Bigg)^{-\frac{1}{3}}
		\nonumber\\
		&-3 \kappa  n^{2} p_{0}^{2}+\kappa  n p_{0}^{2}+3 n-\frac{5}{3}. \nonumber
	\end{align}
\normalsize

From the Equation~\eqref{608} solution and the homogeneous part roots, the general $F(T,B)$ solution in terms of Equations~\eqref{403d} is:
\begin{align}\label{610}
	F(T,B) =& -\Lambda_0+c_1\,T^{\frac{5-9 n}{4}- \frac{\sqrt{81 n^{2}-42 n+9}}{4}}+c_2\,T^{\frac{5-9 n}{4}+ \frac{\sqrt{81 n^{2}-42 n+9}}{4}}+c_3\,T^{-r_p(n)/2}
	\nonumber\\
	&+c_4\, B^{\frac{5-9 n}{4}- \frac{\sqrt{81 n^{2}-42 n+9}}{4}}+c_5\, B^{\frac{5-9 n}{4}+ \frac{\sqrt{81 n^{2}-42 n+9}}{4}}+c_6\, B^{-r_p(n)/2},
\end{align}
where $r_p(n)$ is Equation~\eqref{608}. As in several papers, we find for Equation~\eqref{610} the $F(T,B)=F_1(T)+F_2(B)$ solution form as in refs.~\cite{leonftbcosmo1,leonftbcosmo2,leon2,leon3}. In terms of Equation~\eqref{403e}, the general $F(T,B)$ solution will be:
\small
\vspace{-12pt} 
	\begin{align}\label{610a}
		F(T,B) =& -\Lambda_0+c_1\,\left(T+b_2\,B\right)^{\frac{5-9 n}{4}- \frac{\sqrt{81 n^{2}-42 n+9}}{4}}+c_2\,\left(T+b_2\,B\right)^{\frac{5-9 n}{4}+ \frac{\sqrt{81 n^{2}-42 n+9}}{4}}+c_3\,\left(T+b_2\,B\right)^{-r_p(n)/2},
	\end{align}
\normalsize
where $r_p(n)$ is still described by Equation~\eqref{608}. The teleparallel $F(T,B)$ solutions of Equations~\eqref{610} and~\eqref{610a} are new. {Equations~\eqref{610} and~\eqref{610a} have some similar terms with recent solutions and models in the literature with scalar field source~\cite{leonftbcosmo1,leonftbcosmo2,leon2,leon3,dixit,ftphicosmo}. We note by comparison with the solutions of Sections~\ref{sect41} and~\ref{sect51} that the cosmological constant $-\Lambda_0$ term is present for any other parameter values. This feature contrasts with the perfect fluid solutions and also shows the importance of the cosmological constant in $F(T,B)$ solutions involving DE. We will then be able to compare the new $F(T,B)$ solutions with observational data by using the redshift and Hubble parameter measurement and then applying the torsion-to-scalar ratio technique for the new solutions as done for some perfect fluid $F(T,B)$ solutions in refs.~\cite{dixit,ftbcosmo3}. By this method, we will be able to better elaborate and/or at least improve the DE models.}


\subsubsection{Exponential Scalar Field Solutions}

By setting as scalar field $\phi(t)=p_0\,\exp \left(p\,t\right)$, Equation~\eqref{605} becomes:
\begin{align}\label{607}
	\kappa\,p_0^2\,p^2\,\exp\left(2pt\right)=& \,\frac{\left[-t^3\,\dddot{F}+(9n-8)\,t^2\,\ddot{F}+(21n-10)\,t\,\dot{F}\right]}{36n\left(n-\frac{1}{3}\right)} .  
\end{align}

By setting $p_0=0$ and $F(t)=F_0\,t^r$ ansatz, Equation~\eqref{607} leads to the same solution as Equation~\eqref{606}. For the $p_0 \neq 0$ case (source-based), we will use the modified $F(t)=F_0\,t^r\,\exp(2p\,t)$ ansatz and Equation~\eqref{607} becomes:
\small
\begin{align}\label{609}
	36\kappa_{eff} p^2 p_0^2 n\left(n - \frac{1}{3}\right)=& r\,\left[(r-1)(r-2)+(9n-8)(r-1)+(21n-10)\right]\,t^r
	+2p\Bigg[3r(r-1)
	\nonumber\\
	&+2r(9n-8)+(21n-10)\Bigg]\, t^{r+1}+4p^2\left(3r+9n-8\right)\,t^{r+2}+ (8p^2)\,t^{r+3} ,
\end{align}
\normalsize
where $\kappa_{eff}=\frac{\kappa}{F_0}$. However, there is no exact and direct solution for Equation~\eqref{609}, but we can use the weak approximation $p=\Delta p \approx 0$ limit to solve this last equation. In this case, the homogeneous solution will dominate and the particular part will be considered as a small correction (weak $\phi(t)$ field). By developing up to the $2$nd order in $\Delta p$, Equation~\eqref{609} becomes:
\begin{align}\label{611}
	0\approx & \left(-\frac{r^{3}}{18}+\frac{\left(9 n-5\right) r^{2}}{18}+\frac{2 \left(3 n-1\right) r}{9}\right)+2 t \left(-\frac{r^{2}}{6}+\left(n-\frac{13}{18}\right) r+\frac{7 n}{6}-\frac{5}{9}\right) \Delta p
	\nonumber\\
	&\quad \left(-2 n \kappa  \left(n-\frac{1}{3}\right) p_{0}^{2} t^{-r}+2 \left(n-\frac{r}{3}-\frac{8}{9}\right) t^{2}\right) \Delta p^{2} .
\end{align}

The first part of Equation~\eqref{611} will lead to the same homogeneous parts as in Equation~\eqref{606}. However, the $1$st-order term yields as roots $r=3 n-\frac{13}{6}+\frac{\sqrt{324 n^{2}-216 n+49}}{6}$ and $3 n-\frac{13}{6}-\frac{\sqrt{324 n^{2}-216 n+49}}{6}$. Therefore, the $2$nd-order term takes into account the source terms and leads to $r=-2$ as {the} only possible root. From there, Equation~\eqref{611} will yield as $F(T,B)$ solution in terms of Equations~\eqref{403d}:
\small
\begin{align}\label{613}
	F(T,B) \approx &  -\Lambda_0+c_1\,T^{\frac{5-9 n}{4}- \frac{\sqrt{81 n^{2}-42 n+9}}{4}}+c_2\,T^{\frac{5-9 n}{4}+ \frac{\sqrt{81 n^{2}-42 n+9}}{4}} + \Bigg(c_3\,T^{\frac{7}{12}-\frac{3 n}{2}-\frac{\sqrt{324 n^{2}-216 n+49}}{12}} 
	\nonumber\\
	& + c_4\,T^{\frac{7}{12}-\frac{3 n}{2}+\frac{\sqrt{324 n^{2}-216 n+49}}{12}}\Bigg)\,\left(2\Delta p\right)  +c_5\,B^{\frac{5-9 n}{4}- \frac{\sqrt{81 n^{2}-42 n+9}}{4}}+c_6\,B^{\frac{5-9 n}{4}+ \frac{\sqrt{81 n^{2}-42 n+9}}{4}} 
	\nonumber\\
	&+ \left(c_7\,B^{\frac{7}{12}-\frac{3 n}{2}-\frac{\sqrt{324 n^{2}-216 n+49}}{12}} + c_8\,B^{\frac{7}{12}-\frac{3 n}{2}+\frac{\sqrt{324 n^{2}-216 n+49}}{12}}\right)\,\left(2\Delta p\right) +c_9\,\left(\Delta p \,\right)^2.
\end{align}
\normalsize

The source-based terms at the $2$nd-order approximation lead to a constant contribution to the $F(T,B)$ solution. As in several papers, we find that Equation~\eqref{613} satisfies the    {$F(T,B)=F_1(T)+F_2(B)$} form as in refs.~\cite{leonftbcosmo1,leonftbcosmo2,leon2,leon3,ftphicosmo}. In terms of Equation~\eqref{403e}, the $F(T,B)$ solution will be:
\small
\begin{align}\label{613a}
	F(T,B) \approx &  -\Lambda_0 +c_1\,\left(\Delta p \,\right)^2+c_2\,\left(T+b_2\,B\right)^{\frac{5-9 n}{4}- \frac{\sqrt{81 n^{2}-42 n+9}}{4}}+c_3\,\left(T+b_2\,B\right)^{\frac{5-9 n}{4}+ \frac{\sqrt{81 n^{2}-42 n+9}}{4}} 
	\nonumber\\
	&+ \Bigg(c_4\,\left(T+b_2\,B\right)^{\frac{7}{12}-\frac{3 n}{2}-\frac{\sqrt{324 n^{2}-216 n+49}}{12}} + c_5\,\left(T+b_2\,B\right)^{\frac{7}{12}-\frac{3 n}{2}+\frac{\sqrt{324 n^{2}-216 n+49}}{12}}\Bigg)\,\left(2\Delta p\right) .
\end{align}
\normalsize

 {The previous teleparallel $F(T,B)$ solutions are really new {and also comparable with some recent papers in the literature~\cite{leonftbcosmo1,leonftbcosmo2,leon2,leon3,ftphicosmo}. For the same reasons as in Equations~\eqref{610} and~\eqref{610a},} the cosmological constant is unconditionally present in Equations~\eqref{613} and~\eqref{613a} and this shows the necessity of this term in DE models. Here again, we will be able to use observational redshift and Hubble parameter data, and then the tensor-to-scalar methods to compare the new $F(T,B)$ solutions as for the previous cases, as performed in recent papers~\cite{dixit,ftbcosmo3}.}


\subsection{$k=+1$ Case}\label{sect63}

By merging Equations~\eqref{464a}--\eqref{464b} together, performing $F(t)=F_0\,t^r$ (power-law scalar field {cases}) and then $F(t)=F_0\,t^r\,\exp\left(2\Delta p\,t\right)$ ansatz (where $\Delta p\,\rightarrow\,0$ limit and for exponential scalar field {cases}), we find as a solution for most values of $n$ the cosmological constant $F(T,B)=-\Lambda_0$ solution and a constant scalar field source $\phi=p_0$. But there are, for specific values of $n$, some new teleparallel $F(T,B)$ solutions, especially in   {terms of Equation~\eqref{462d}.}

\subsubsection{Power-Law Scalar Field Solutions}

\begin{enumerate} 
	\item {${\bf n\rightarrow\,\frac{1}{3}}$}: The unified Equations~\eqref{464a}--\eqref{464b} for $n=\frac{1}{3}+\epsilon$ and $p_0=0$ leads to $r=0$ and $r=-2$. For $p_0\neq 0$ (source-based), we set $p=\frac{r}{2}+1$ in the unified Equations~\eqref{464a}--\eqref{464b} and we find that $r \approx -2+\mathcal{O}\left(\epsilon\right)$ at the 1st order in $\epsilon$. In terms of Equation~\eqref{462d}, the general $F(T,B)$ solution is:
	\begin{align}\label{680}
		F(T,B) =& -\Lambda_0 +c_1\,\left(\frac{2B}{3\epsilon}-T\right)^{3}+\mathcal{O}\left( \left(\frac{2B}{3\epsilon}-T\right)^{3-\mathcal{O}\left(3\epsilon/2\right)}\right).
	\end{align}

	\item {${\bf n=1}$}: The unified Equations~\eqref{464a}--\eqref{464b} become, by setting $p=\frac{r}{2}+1$ and   {for $a_0^2 \neq 1$:}
	
	\vspace{-12pt} 
		\begin{align}\label{681}
			0=&  \left(a_{0}^{2}-1\right) r^{3}+2 \left(3 p_{0}^{2} \kappa  \left(a_{0}^{2}-1\right)-2 a_{0}^{2}\right) r^{2}-8 \left(a_{0}^{2}-1\right) \left(1-3 p_{0}^{2} \kappa \right) r
			+24 p_{0}^{2} \kappa  \left(a_{0}^{2}-1\right) .
		\end{align}
	
	For the $p_0=0$ case, we find that $r=r_{\pm}(a_0)=\frac{2 \left(a_{0}^{2}\pm\sqrt{3 a_{0}^{4}-4 a_{0}^{2}+2}\right)}{a_{0}^{2}-1}$ and $0$ ($-\Lambda_0$ term). By adding the $p_0\neq 0$ (source-based) terms by setting $p=\frac{r}{2}+1$ in Equations~\eqref{464a}--\eqref{464b}, the Equation~\eqref{681} solution in terms of Equation~\eqref{462d} will be:
	\small
	\begin{align}\label{682}
		F(T,B) =&-\Lambda_0+ c_1\,\left(3B-T\right)^{-\frac{r_{+}(a_0)}{2}}+ c_2\, \left(3B-T\right)^{-\frac{r_{-}(a_0)}{2}} +c_3\, \left(3B-T\right)^{-r(a_0,p_0)/2},
	\end{align} 
	\normalsize
	where
	\small
	\vspace{-12pt} 
		\begin{align}\label{683}
			r(a_0,p_0)=& \frac{1}{3   \left(a_{0}^{2}-1\right)} \Bigg(-216 a_{0}^{6} \kappa^{3} p_{0}^{6}+1080   a_{0}^{6} \kappa^{2} p_{0}^{4}+648 a_{0}^{4} \kappa^{3} p_{0}^{6}-1260   a_{0}^{6} \kappa  p_{0}^{2}-2808   a_{0}^{4} \kappa^{2} p_{0}^{4}
			\nonumber\\
			&-648 a_{0}^{2} \kappa^{3} p_{0}^{6}+208   a_{0}^{6}+2772   a_{0}^{4} \kappa  p_{0}^{2}+2376   a_{0}^{2} \kappa^{2} p_{0}^{4}+216 \kappa^{3} p_{0}^{6}-288   a_{0}^{4}
			+12\, \left(a_{0}^{2}-1\right)^2
			\nonumber\\
			&\times\,  \Bigg(-\frac{3}{\left(a_{0}^{2}-1\right)} \Bigg(108 a_{0}^{6} \kappa^{3} p_{0}^{6}-315   a_{0}^{6} \kappa^{2} p_{0}^{4}-108 a_{0}^{4} \kappa^{3} p_{0}^{6}-120   a_{0}^{6} \kappa  p_{0}^{2}+9   a_{0}^{4} \kappa^{2} p_{0}^{4}
			\nonumber\\
			&-108 a_{0}^{2} \kappa^{3} p_{0}^{6}+48   a_{0}^{6}+624   a_{0}^{4} \kappa  p_{0}^{2}+639   a_{0}^{2} \kappa^{2} p_{0}^{4}+108 \kappa^{3} p_{0}^{6}-112   a_{0}^{4}-696   a_{0}^{2} \kappa  p_{0}^{2}
			\nonumber\\
			&-333   \kappa^{2} p_{0}^{4}+96   a_{0}^{2}+288 p_{0}^{2} \kappa   -32  \Bigg)\Bigg)^{1/2}-2052   a_{0}^{2} \kappa  p_{0}^{2}-648   \kappa^{2} p_{0}^{4}+144   a_{0}^{2}+540 p_{0}^{2} \kappa   \Bigg)^{\frac{1}{3}}
			\nonumber\\
			&+\frac{4 \left(9 a_{0}^{4} \kappa^{2} p_{0}^{4}-30   a_{0}^{4} \kappa  p_{0}^{2}-18 a_{0}^{2} \kappa^{2} p_{0}^{4}+10   a_{0}^{4}+48   a_{0}^{2} \kappa  p_{0}^{2}+9 \kappa^{2} p_{0}^{4}-12   a_{0}^{2}-18   \kappa  p_{0}^{2}+6  \right)}{3   \left(a_{0}^{2}-1\right)}
			\\
			& \times \Bigg(-216 a_{0}^{6} \kappa^{3} p_{0}^{6}+1080   a_{0}^{6} \kappa^{2} p_{0}^{4}+648 a_{0}^{4} \kappa^{3} p_{0}^{6}-1260   a_{0}^{6} \kappa  p_{0}^{2}-2808   a_{0}^{4} \kappa^{2} p_{0}^{4}-648 a_{0}^{2} \kappa^{3} p_{0}^{6}
			\nonumber\\
			&+208   a_{0}^{6}+2772   a_{0}^{4} \kappa  p_{0}^{2}+2376   a_{0}^{2} \kappa^{2} p_{0}^{4}+216 \kappa^{3} p_{0}^{6}-288   a_{0}^{4}+12\, \left(a_{0}^{2}-1\right)^2  \Bigg(-\frac{3 }{\left(a_{0}^{2}-1\right)} 
			\nonumber\\
			&\times\,\Bigg(108 a_{0}^{6} \kappa^{3} p_{0}^{6}-315   a_{0}^{6} \kappa^{2} p_{0}^{4}-108 a_{0}^{4} \kappa^{3} p_{0}^{6}-120   a_{0}^{6} \kappa  p_{0}^{2}+9   a_{0}^{4} \kappa^{2} p_{0}^{4}-108 a_{0}^{2} \kappa^{3} p_{0}^{6}+48 a_{0}^{6}
			\nonumber\\
			&+624   a_{0}^{4} \kappa  p_{0}^{2}+639   a_{0}^{2} \kappa^{2} p_{0}^{4}+108 \kappa^{3} p_{0}^{6}-112   a_{0}^{4}-696   a_{0}^{2} \kappa  p_{0}^{2}-333   \kappa^{2} p_{0}^{4}+96   a_{0}^{2}+288 p_{0}^{2} \kappa  
			\nonumber\\
			& -32  \Bigg)\Bigg)^{1/2} -2052   a_{0}^{2} \kappa  p_{0}^{2}-648   \kappa^{2} p_{0}^{4}+144   a_{0}^{2}
			+540 p_{0}^{2} \kappa   \Bigg)^{-\frac{1}{3}} 
			\nonumber\\
			&  +\frac{2 \left(-3 p_{0}^{2} \kappa  a_{0}^{2}+2   a_{0}^{2}+3 p_{0}^{2} \kappa \right)}{3   \left(a_{0}^{2}-1\right)}.\nonumber
		\end{align}
	\normalsize
	
\end{enumerate}

{Equations~\eqref{680} and~\eqref{682} are new teleparallel $F(T,B)$ solutions and no direct equivalent exists in the recent literature. However, we are also able to compare these solutions with observational data by using the fitting techniques in refs.~\cite{dixit,ftbcosmo3} as previously.}

\subsubsection{Exponential Scalar Field Solutions}

By setting $p=\Delta p \rightarrow\,0$, we find at the $1$st order the unified Equations~\eqref{464a}--\eqref{464b}:
\begin{enumerate} 
	\item {${\bf n\rightarrow\,\frac{1}{3}}$}:
	\begin{align}\label{684}
		0=& r^{2} \left(r+2\right)\Bigg(-a_{0}^{4}  t^{\frac{10}{3}}+2 a_{0}^{2}t^{\frac{14}{3}}-\frac{a_{0}^{8}}{54} t^{\frac{2}{3}}-\frac{3}{2} t^{2} \left(-\frac{4 a_{0}^{6}}{27}+t^{4}\right)\Bigg)
		\nonumber\\
		&\quad +\left(1+r^{2}+\frac{7}{3} r\right)\Bigg(-6 a_{0}^{4} t^{\frac{13}{3}}+12  a_{0}^{2} t^{\frac{17}{3}}-\frac{a_{0}^{8} t^{\frac{5}{3}}}{9}-9 t^{3} \left(-\frac{4 a_{0}^{6}}{27}+t^{4}\right)\Bigg) \Delta p ,
	\end{align}
	where $\epsilon=0$. Equation~\eqref{684}'s dominating terms' roots are $r=0$ and $-2$. For the $1$st-order term, Equation~\eqref{684} roots are $r_{\pm}=-\frac{\left(7\pm\sqrt{13}\right)}{6}$ and the $2$nd-order term leads by default to $r=-2$ with the $p_0^2=\frac{7 a_{0}^{2}-3}{6 \kappa  \left(a_{0}^2-1\right)}$ condition. In terms of Equation~\eqref{462d}, the Equation~\eqref{684} solution is:
	\small
	 {\begin{align}\label{685}
			F(T,B)\approx  & -\Lambda_0 + c_1\,\left(\frac{2B}{3\epsilon}-T\right)^{3}+\Bigg[c_2\,\left(\frac{2B}{3\epsilon}-T\right)^{\frac{\left(1+\sqrt{13}\right)}{4}}+c_3\,\left(\frac{2B}{3\epsilon}-T\right)^{\frac{\left(1-\sqrt{13}\right)}{4}}\Bigg]\,\left(2\Delta p\right)
			\nonumber\\
			&\;+ 2c_4\,\left(\Delta p\right)^2.
	\end{align}}
	\normalsize
	

	\item {${\bf n=1}$}: 
	\vspace{-12pt} 
		\begin{align}\label{686}
			0=& \left(-a_{0}^{2} r^{3}+4 a_{0}^{2} r^{2}+8 a_{0}^{2} r+r^{3}-8 r\right) -6 \left(a_{0}^{2} r^{2}-\frac{5}{3} a_{0}^{2} r-r^{2}-\frac{11}{3} a_{0}^{2}-r+\frac{7}{3}\right) t \Delta p .
		\end{align}
	
	Equation~\eqref{686}'s dominating part leads to $r=r_{\pm}=\frac{2 \left(a_{0}^{2}\pm\sqrt{3 a_{0}^{4}-4 a_{0}^{2}+2}\right)}{a_{0}^{2}-1}$ and $0$. The $1$st-order terms lead to $r_{\pm}=\frac{5 a_{0}^{2}+3\pm\sqrt{157 a_{0}^{4}-186 a_{0}^{2}+93}}{6 \left(a_{0}^{2}-1\right)}$ and the $2$nd-order terms to $r=-2$ under the $p_0^2=\frac{7 a_{0}^{2}-3}{6 \kappa  \left(a_{0}^2-1\right)}$ condition. The Equation~\eqref{686} solution in terms of Equation~\eqref{462d} is:
	\small
	\begin{align}\label{687}
		F(T,B)\approx & -\Lambda_0 + 2c_1\,\left(\Delta p\right)^{2}+ c_2\,\left(3B-T\right)^{-\frac{\left(a_{0}^{2}+\sqrt{3 a_{0}^{4}-4 a_{0}^{2}+2}\right)}{\left(a_{0}^{2}-1\right)}}+c_3\,\left(3B-T\right)^{-\frac{\left(a_{0}^{2}-\sqrt{3 a_{0}^{4}-4 a_{0}^{2}+2}\right)}{\left(a_{0}^{2}-1\right)}}
		\nonumber\\
		&+\Bigg[c_4\,\left(3B-T\right)^{-\frac{11a_{0}^{2}-3+\sqrt{157 a_{0}^{4}-186 a_{0}^{2}+93}}{12 \left(a_{0}^{2}-1\right)}}+c_5\,\left(3B-T\right)^{-\frac{11a_{0}^{2}-3-\sqrt{157 a_{0}^{4}-186 a_{0}^{2}+93}}{12 \left(a_{0}^{2}-1\right)}}\Bigg]\,\left(2\Delta p\right)  .
	\end{align}
	\normalsize

	\item  {${\bf n\gg 1}$}: The approximated and unified Equations~\eqref{464a}--\eqref{464b} (keeping the highest power-terms) are:
	\begin{align}\label{688}
		0\approx &  r \left(r^{2}-175 r-236\right) +6 \left(r^{2}-\frac{347}{3} r-\frac{410}{3}\right)\,t \Delta p.
	\end{align}
	
	Equation~\eqref{688}'s dominating terms' roots are $r=r_{\pm}=\frac{175}{2}\pm\frac{\sqrt{31,569}}{2}$ and $0$. For the $1$st-order term, the roots are $\frac{347}{6}\pm\frac{\sqrt{125,329}}{6}$ and the $2$nd-order term leads again to $r=-2$ and $p_0^2=\frac{89}{1770 \kappa}$ condition. The Equation~\eqref{688} solution in terms of Equation~\eqref{462d} is:
	\small
	\begin{align}\label{689}
		F(T,B) \approx & -\Lambda_0+2c_1\,\left(\Delta p\right)^2+c_2\,\left(2B-T\right)^{-\frac{175}{80}-\frac{\sqrt{31,569}}{80}}+c_3\,\left(2B-T\right)^{-\frac{175}{80}+\frac{\sqrt{31,569}}{80}}
		\nonumber\\
		&+\Bigg[c_4\,\left(2B-T\right)^{-\frac{353}{240}-\frac{\sqrt{125,329}}{240}}+c_5\,\left(2B-T\right)^{-\frac{353}{240}+\frac{\sqrt{125,329}}{240}}\Bigg]\left(2\Delta p\right).
	\end{align}
	\normalsize
	
\end{enumerate}

{The new teleparallel $F(T,B)$ solutions expressed in Equations~\eqref{685},~\eqref{687} and~\eqref{689} have no real equivalents compared to the recent papers in the literature. These new solutions can be compared in the future with observational data by the fitting techniques in refs.~\cite{dixit,ftbcosmo3} as usual.}

\subsection{$k=-1$ Case}\label{sect62}

By putting together Equations~\eqref{434a}--\eqref{434b} and performing $F(t)=F_0\,t^r$ (power-law scalar field), and then $F(t)=F_0\,t^r\,\exp\left(2\Delta p\,t\right)$ ansatz (where $\Delta\,p\rightarrow\,0$ limit and for exponential scalar field), we again find as a solution for most values of $n$ the cosmological constant $F(T,B)=-\Lambda_0$ solution and a constant scalar field source $\phi=p_0$. But there are, for specific values of $n$, new teleparallel $F(T,B)$ solutions and specifically in terms of Equation~\eqref{432d}.

\subsubsection{Power-Law Scalar Field Solutions}

\begin{enumerate} 
	\item {${\bf n=1}$}: The unified Equations~\eqref{434a}--\eqref{434b}, by setting $p=\frac{r}{2}+1$, yield as roots $r=0$ and $-2$ for $p_0=0$ . For $p_0\neq 0$ (source-based), the root $r=-2$ is confirmed, leading to a constant scalar field ($p=0$) and the $F(T,B)$ solution in terms of Equation~\eqref{432d} is exactly Equation~\eqref{490} found in Section~\ref{sect42}.
	
	\item {{${\bf n\gg 1}$}: By keeping only the leading terms, we find unified Equations~\eqref{434a}--\eqref{434b} and the only $\mathbb{R}$-valued root:}
	\small
	
		\begin{align}\label{690}
			0=& 17,405  r^{3}+\frac{1}{8}\left(492,909,600 \kappa  p_{0}^{2}-5,151,880  \right) r^{2}+\frac{1}{8}\left(1,971,638,400 \kappa  p_{0}^{2}+5,569,600  \right) r
			\nonumber\\
			& \quad +246,454,800 \kappa  p_{0}^{2} ,
			\nonumber\\
			\Rightarrow &\,r(p_0)\approx \frac{1}{3}\Bigg(-44,361,864,000 \kappa^{3} p_{0}^{6}+1,616,576,400   \kappa^{2} p_{0}^{4}-16,450,380 \kappa  p_{0}^{2}  +43,993  
			\nonumber\\
			&\quad\quad\quad\quad+36 \sqrt{-5 \left(43,622,499,600 \kappa^{3} p_{0}^{6}-1,067,483,460   \kappa^{2} p_{0}^{4}+3,491,207 \kappa  p_{0}^{2}  +2015  \right)}\,  \Bigg)^{\frac{1}{3}}
			\nonumber\\
			&\quad\quad\quad\quad+\frac{-3540 \kappa  p_{0}^{2}+37}{3}	+\frac{12,531,600 \kappa^{2} p_{0}^{4}-304,440   \kappa  p_{0}^{2}+1249}{3} \Bigg(-44,361,864,000 \kappa^{3} p_{0}^{6}
			\nonumber\\
			&\quad\quad\quad\quad+1,616,576,400   \kappa^{2} p_{0}^{4}-16,450,380 \kappa  p_{0}^{2}  +43993  
			\nonumber\\
			&\quad\quad\quad\quad+36 \sqrt{-5 \left(43,622,499,600 \kappa^{3} p_{0}^{6}-1,067,483,460   \kappa^{2} p_{0}^{4}+3,491,207 \kappa  p_{0}^{2}  +2015  \right)}\Bigg)^{-\frac{1}{3}}.
		\end{align}
	\normalsize
	The Equation~\eqref{690} solution in terms of Equation~\eqref{432d} under the $n\rightarrow\,\infty$ limit is:
	\begin{align}\label{691}
		F(T,B) \approx & -\Lambda_0 + c_1\,\left(\frac{\sqrt{T}+\delta_1\sqrt{T+B}}{B} \right)^{r(p_0)}.
	\end{align}
	
\end{enumerate}

{Equation~\eqref{691} comprises new teleparallel $F(T,B)$ solutions and no direct equivalent exists in recent papers. This last $F(T,B)$ solution can be compared with observational data from the fitting techniques in refs.~\cite{dixit,ftbcosmo3}, as for previous new solutions.}

\subsubsection{Exponential Scalar Field Solutions}


\begin{enumerate} 
	\item {${\bf n=1}$:} The only possible unified solutions of Equations~\eqref{434a}--\eqref{434b} from the dominating terms with $\delta=1$ are $r=0$ and $-2$ roots. The $1$st-order terms in $\Delta p$ with $\delta=1$ are $r=-1$ and $-3$ roots, and the $F(T,B)$ solution in terms of Equation~\eqref{432d} will be:
	\begin{align}\label{692}
		F(T,B) =& -\Lambda_0+2c_1\Delta p +c_2\,\left(\frac{B^2}{T}\right)\left(1+2c_3 \Delta p\right). 
	\end{align}
	The $2$nd-order term leads to $p_0 \rightarrow 0$. Equation~\eqref{692} has the same form as Equation~\eqref{490} and the only differences are the shifted coefficients. The exponential scalar field will shift the coefficient including the cosmological constant, as we can see in   {Equation~\eqref{692}}. For $\Delta p=0$, we obtain exactly Equation~\eqref{490}.
	
	\item  {${\bf n\gg 1}$:} For the unified Equations~\eqref{434a}--\eqref{434b}, we find as roots $r=0$ for the dominating term and $r=-1$ for the $1$st-order term in $\Delta p$. We can neglect the $2$nd-order parts and the solution is only the shifted cosmological constant $F(T,B)= -\Lambda_0+2c_1\,\Delta p$, a GR solution. Once again, the exponential scalar field will fundamentally shift the cosmological constant as in Equation~\eqref{692}.
	
\end{enumerate}

{Once again, there is no equivalent $F(T,B)$ solution of Equation~\eqref{692} in the recent literature. But we can still perform the observational data comparison techniques as proposed in refs.~\cite{dixit,ftbcosmo3}.}


\section{Discussion and Conclusions}\label{sect7}

We found for the flat cosmological ($k=0$) cases some exact and general teleparallel $F(T,B)$ solutions satisfying the $F_1(T)+F_2(B)$ form. This situation is possible because the time coordinate can be separately expressed in terms of torsion scalar $T$ or the boundary variable $B$ as shown in Equations~\eqref{403d}. This situation has the advantage of separately studying the pure $F(T)$ and $F(B)$ flat cosmological model solution in teleparallel gravity for any type of source: perfect fluids as scalar fields. This fact has been proven in this paper, especially with the TRW $F(T,B)$ generalization solutions for {the linear and non-linear perfect fluid cases.} This {feature} also shows at least for TRW spacetimes that the teleparallel $F(T,B)$-gravity theory is really an extension of the pure teleparallel $F(T)$-gravity by keeping a very similar symmetry and frame approach. However, Equation~\eqref{403d} for $t$ in terms of $T$ and $B$ is not a unique method yielding $F(T,B)$ solutions. On this point, Equation~\eqref{403e} expresses the most general method to obtain a great number of new additional teleparallel $F(T,B)$ solutions. Even if the recent literature led to similar $F(T,B)$ solution forms, we have used a more fundamental approach to find the most general and specific possible solutions in this work. Nevertheless, the teleparallel $F_1(T)+F_2(B)$ class of solutions is useful for separately studying the impacts and contributions of $T$-based and $B$-based terms in future works on teleparallel cosmological models. {Furthermore, we will be able to study the physical impacts and contributions of each of variable $T$ and $B$ separately on the DE processes by using the most general $F(T,B)$ solutions of the current research work compared to recent papers~\cite{cosmobaha,dixit,cosmoftbmodels,cosmoftbenergy,ftbcosmo2,ftbcosmo3,ftbcosmo4,ftbcosmo5,ftbcosmo6,ftphicosmo}.}

For the $k=-1$ and $+1$ cases, we only found teleparallel $F(T,B)$ solutions for well-specified and characteristic $n$ values. As stated along this development, the cosmological constant (i.e., $F(T,B)=-\Lambda_0$) is the default solution for most of $n$ the values, because Equations~\eqref{464a}--\eqref{464b} and~\eqref{434a}--\eqref{434b} are fundamentally difficult to solve. {However, the cosmological constant $\Lambda_0$ solution basically states in a physically sound manner that DE is dominating in the universe.} It was a challenge to find specific cases in which the FEs yield new analytical $F(T,B)$ solutions, but we achieved {that aim} in this research work. We even found for $k=-1$ and $n=1$ subcase situations Equation~\eqref{490} (and Equation~\eqref{692} as a generalization for exponential scalar field) arising for perfect fluids as for scalar field sources. This particular case deserves more investigation concerning the physical processes underlying this specific teleparallel $F(T,B)$ solution. For the $k=+1$ cases, we have obtained solutions for $n=\frac{1}{3}$, $1$ and the large $n$ limit (e.g., $n=20$ and larger). Even if Equations~\eqref{464a}--\eqref{464b} are less difficult to solve, the number of possible $k=+1$ solution cases is quite limited. The $F(T,B)$ solutions are essentially characterized by $(b_2\,B+T)^q$ terms and/or their superposition, but the physical process and the explanations behind this proportionality also deserve to be investigated in future works on teleparallel cosmological models. {Indeed, we can expect that the lower values of $n$ should be more useful for DE quintessence and/or bouncing physical models and higher values of $n$ for phantom energy physical models.}

We achieved this primary aim by solving and finding the exact teleparallel $F(T,B)$ solutions for linear, quadratic perfect fluids as for power-law and exponential scalar field sources. {Primarily, a first apparent point concerns the large number of exact $k=0$ case $F(T,B)$ solutions compared to the new and possible $k=\pm 1$ cases' exact $F(T,B)$ solutions. Out of the FE explanations, the literature on the subject greatly highlights this disparity between cosmological $F(T,B)$ solution classes. In the present work, we have greatly reduced this disparity by finding an important number of new $F(T,B)$ solutions for the $k=\pm 1$ cases while we find new additional $F(T,B)$ solutions for the $k=0$ cases. In specific cases, there are generalizations of results from the recent literature as mentioned in the previous sections.

	Concerning the results in terms of gravitational sources,} we found several common points between the new teleparallel $F(T,B)$ solutions for perfect fluid and scalar field sources, allowing the further study of the fundamental quintessence process {and also some bouncing models.} Another important and common point arising in Section~\ref{sect6} is that all new $F(T,B)$ solutions are valid for any scalar potential source $V(\phi)$, which will be very helpful for cosmological studies. The similar terms and the scalar potential invariance for a large number of new $F(T,B)$ solutions will facilitate the future and detailed DE quintessence model studies by making parallels between the new teleparallel $F(T,B)$ solutions found in the current research investigation{~\cite{cosmoftbmodels,steinhardt1,steinhardt2,cosmobaha,cosmoftbenergy,ftbcosmo2,ftbcosmo3,ftbcosmo4,ftbcosmo5,ftbcosmo6,ftphicosmo}. This potential independence of $F(T,B)$ solutions will allow work with any universe models involving DE, regardless of the $V(\phi)$ where the physical system is evolving.} In addition, some of the new $F(T,B)$ solutions will also be useful for future studies on teleparallel phantom energy models. For this purpose, the new $F(T,B)$ solutions for the large $n$ limit case are relevant as a concrete fast universe expansion scenario model and may deserve some future studies on phantom energy models~\cite{cosmoftbmodels,caldwell1,farnes,baumframpton}. It is known that a fast and uncontrolled universe expansion scenario will lead to the Big Rip event at the end, according to this extreme scenario model. All these possible studies greatly deserve to be achieved in the near future within a teleparallel $F(T,B)$-type framework {and the current investigation provides the relevant and up-to-date $F(T,B)$ solutions for this ultimate aim.

	However, these new types of possible physical model studies will have to be accompanied by more specific comparisons of the new and fresh teleparallel $F(T,B)$ solutions found here with observation data, in particular by using the redshift measurement data~\cite{dixit,ftbcosmo3}. This sort of more comparative study is usually carried out more precisely by studying the effects on Hubble parameters and other relevant physical quantities such as effective densities and pressures~\cite{dixit,ftbcosmo3}. In this direction, the same studies will need to fully address the other thermodynamic effects such as the entropy and energy conditions related with the different types of DE models as expected. Ideally, this type of strong-detailed study will need to be carried out for each individual new $F(T,B)$ solution and would have largely exceeded the scope of this current paper, whose aims were much more theoretical and mathematical. In this sense, the new results found in the current paper have been compared to other recent theoretical-based studies involving several types of $F(T,B)$ solutions, which is already a good achievement. But an observational data-based study easily deserves a logical continuation of the present research work.}

In some recent work on cosmological teleparallel gravity models, there are some papers solving for perfect fluid sources in more refined spacetimes in teleparallel $F(T)$-type gravity, such as the Kantowski--Sachs spacetimes~\cite{nonvacKSpaper,roberthudsonSSpaper}. There is also a {new submitted research work on teleparallel $F(T)$ solutions in the} Kantowski--Sachs spacetimes with a scalar field source~\cite{scalarfieldKS}. {In addition, there is another paper in final preparation on TRW spacetime $F(T)$-type solutions with similar scalar field sources to complete the possible new teleparallel solution finding~\cite{scalarfieldTRW}.} This type of work not only aims to find new teleparallel $F(T)$ solutions useful for DE quintessence and phantom energy with Big Rip processes, but it may also offer some possible generalizations to the teleparallel   {$F(T,B)$-gravity} extension for spacetime curvature inclusion into these more developed models. Indeed, there is an important number of possible teleparallel extensions in the literature which may be applied to more complex cosmological models useful for complete {dark energy} quintessence, {phantom energy and quintom} theories. These extensions can go further than simple TRW spacetimes. These possibilities should be considered and investigated seriously in the future.

\vspace{6pt}

\section*{Acknowledgements}

AL is supported by an Atlantic Association of Research in Mathematical Sciences (AARMS) fellowship. Thanks to Coley A.A. and van den Hoogen R.J. for helpful discussion and their opinion on the paper.








\section*{Abbreviations}
{
	The following abbreviations are used in this manuscript:\\
	
	\noindent 
	\begin{tabular}{@{}ll}
		A.L. & Alexandre Landry \\
		{DE} & {Dark Energy} \\
		EoS & Equation of State\\
		Eqn & Equation \\
		FE & Field Equation\\
		F.G. & Fateme Gholami \\
		GR & General Relativity \\
		KG & Klein-Gordon\\
		KV & Killing Vector\\
		Ref. & Reference\\
		TRW & Teleparallel Robertson-Walker\\
		
	\end{tabular}
}

\appendix
\section[\appendixname~\thesection]{Power-Law Ansatz Field Equations}
\subsection[\appendixname~\thesubsection]{$k=+1$ Case}\label{appen1b}

Equations~\eqref{361a}--\eqref{361b} becomes for any fluid of EoS $P=P(\rho)$:
\small

\vspace{-12pt} 
	\begin{align}
		\kappa\,\rho =& \, -\frac{F}{2}+\frac{\left[t^2\,\ddot{F}+4\,t\,\dot{F}\right]}{12\left(n-\frac{1}{3}\right)}-\frac{n}{4\left(n-\frac{1}{3}\right)}\,\frac{\left[3a_0^2\left(n-\frac{2}{9}\right)\,t^{2n-2}-1\right]}{\left(n a_0^2\,t^{2n-2}-1\right)}\,t\,\dot{F} ,  \label{464a}
		\\
		\kappa\left(\rho+3P(\rho)\right) =& \,\frac{1}{2\left(n a_0^2\,t^{2n-2}-1\right)^2} \Bigg[-2a_0^2\, \left(n+\frac{1}{2}\right)\,t^{2n-1}\,\dot{F}+3a_0^4\,n\,t^{4n-3}\,\dot{F}-a_0^2\,t^{2n}\,\ddot{F}+a_0^4\,n\,t^{4n-2}\,\ddot{F}
		\nonumber\\
		&\,+2\,\frac{\left(n a_0^2\,t^{2n-2}-1\right)}{n\left(n-\frac{1}{3}\right)}\Bigg(\left(na_0^2\,t^{2n-2}-1\right)\left(-\frac{t^3\,\dddot{F}}{12}+\frac{n-6}{12}\,t^2\,\ddot{F}+n\left(n-\frac{1}{3}\right)F\right)
		\nonumber\\
		&+\frac{3na_0^2}{2}\left(n^2-\frac{n}{2}-\frac{2}{9}\right)\,t^{2n-1}\,\dot{F}-\frac{1}{2}\,\left(n^2+\frac{7n}{6}-\frac{4}{3}\right)\,t\,\dot{F}\Bigg)\Bigg] ,  \label{464b}
	\end{align}
\normalsize
{where $n\neq \left\lbrace 0,\,\frac{1}{3}\right\rbrace$, $F=F(t)$, $\dot{F}=F_t$ and $t$ is the time coordinate. We will find}    the Equations~\eqref{464a}--\eqref{464b} for various cases, especially by applying the $F(t)=F_0\,t^r$ power-law ansatz.

\subsection[\appendixname~\thesubsection]{$k=-1$ Case}\label{appen1a}

Equations~\eqref{331a}--\eqref{331b} becomes for any fluid of EoS $P=P(\rho)$:
\small
\vspace{-12pt} 
	\begin{align}\label{434a}
		\kappa\,\rho =& \,\frac{\left(a_{0} n +\delta\,t^{1-n}\right)}{4 n \left(a_{0} \left(3n -1\right) +\delta  \left(n +1\right)\,t^{1-n}\right)^{2}} \Bigg( t\, \dot{F} \left(3a_{0} \left(3n -1\right)+\delta \left(n +2\right) \left(n +1\right) t^{1-n}\right)
		\nonumber\\
		&\,+t^2\,\ddot{F} \left(a_{0} \left(3n -1\right) +\delta  \left(n +1\right)\,t^{1-n}\right)\Bigg)+\frac{a_0\,t \dot{F}}{4 \left(a_{0} \left(3n -1\right) +\delta  \left(n +1\right)\,t^{1-n}\right)}
		\\
		&\, -\frac{t \dot{F} \left(2a_{0} \left(a_{0} \left(3n -1\right) +\delta  \left(n +1\right)\,t^{1-n}\right)+3 \left(a_{0}+\delta  t^{1-n} \right) \left(n a_{0} +\delta  t^{1-n} \right)\right)}{4 \left(a_{0} +\delta  t^{1-n} \right) \left(a_{0} \left(3n -1\right) +\delta  \left(n +1\right)\,t^{1-n}\right)}-\frac{F}{2} ,\nonumber  
	\end{align}
	\begin{align}\label{434b}
		\kappa\left(\rho+3P(\rho)\right) =& \,F+\frac{t \dot{F} \left(2 \left(a_{0} \left(3 n-1\right) t^{n}+t \delta  \left(n+1\right)\right) a_{0} t^{2 n}+3t^{n} \left(a_{0} t^{n}+\delta  t\right) \left(n a_{0} t^{n}+\delta  t\right)\right) \left(a_{0} n+\delta  \,t^{1-n}\right)}{2 \left(n a_{0} t^{n}+\delta  t\right) \left(a_{0} t^{n}+\delta  t\right) \left(a_{0} \left(3 n-1\right) t^{n}+t \delta  \left(n+1\right)\right)}
		\nonumber\\
		&\,-\frac{\left(a_{0} \delta \left(n+1\right) t^{1-n}+ a_{0}^{2} \left(3n-1\right) +\left(a_{0}+\delta  t^{1-n}\right) \left(n a_{0} +\delta  t^{1-n}\right)\right) a_{0} t\,\dot{F}}{2 \left(\delta\left(n+1\right) t^{1-n}+a_{0} \left(3n-1\right)\right) \left(n a_{0}+\delta  t^{1-n}\right) \left(a_{0}+\delta  t^{1-n}\right)}
		\nonumber\\
		&\,+\Bigg[2a_0 t^{2+6n} \left(n a_{0}+\delta  t^{1-n}\right) \left(a_{0}+\delta  t^{1-n}\right)^{2}\left(a_{0} \left(3 n-1\right) + \delta  \left(n+1\right)t ^{1-n}\right)^{2}\Bigg]^{-1} 
		\nonumber\\
		&\,\times\,\Bigg(\dot{F} \left(n a_{0} t^{n}+\delta  t\right)^{2} \left(a_{0} t^{n}+\delta  t\right)^{2} \left[\frac{3a_{0}^{2}}{2} \left(3n-1\right) t^{2+2n}+ t^{3+n} \frac{a_{0} \delta}{2} \left(n+2\right) \left(n+1\right) \right]
		\nonumber\\
		&\,+\Bigg(\frac{\ddot{F}}{2} \left(a_{0} t^{n}+\delta  t\right)^{2} \left(n a_{0} t^{n}+\delta  t\right)^{2} t^{3+n}+t^{2+2n} a_{0} \left(a_{0} \left(3 n-1\right) t^{n}+t \delta  \left(n+1\right)\right) 
		\nonumber\\
		&\,\times\,\left(t \left(a_{0} t^{n}+\delta  t\right) \left(n a_{0} t^{n}+\delta  t\right) \ddot{F}+\dot{F} \left(3 n a_{0}^{2} t^{2n}+a_{0} \delta  \left(n+2\right) \left(n+1\right) t^{n+1}+\left(2n+1\right) t^{2}\right)\right)\Bigg)
		\\
		&\,\times\, a_{0} \left(a_{0} \left(3 n-1\right) t^{n}+t \delta  \left(n+1\right)\right)\Bigg)+\frac{\dot{F}}{2} \Bigg(-\frac{t a_{0} \left(t^{1-n} \delta  \left(n^{2}+3n+2 \right)+9 a_{0} n-3 a_{0}\right)^{2}}{n \left(\delta  t^{1-n} n+\delta  t^{1-n}+3 a_{0} n-a_{0}\right)^{3}}
		\nonumber\\
		&\,+\frac{a_{0}^{2} \left(\frac{t^{n+2} \delta}{a_{0}} \left(n^{4}+6  n^{3} +11  n^{2} +6  n \right)-12 \,t^{2n+1} n(1-3n)\right)}{2 n^{2} \left(3 n a_{0} t^{n}+\delta  n t-a_{0} t^{n}+\delta  t\right)^{2}}\Bigg)
		\nonumber\\
		&\,-\frac{\ddot{F} \left(3 \left(3n-1\right) a_{0}^{2} t^{2+2 n}+a_{0} \delta  \left(n+2\right) \left(n+1\right) t^{3+n}\right)}{2 n \left(a_{0} \left(3 n-1\right) t^{n}+t \delta  \left(n+1\right)\right)^{2}}-\frac{\dddot{F} t^{3} a_{0}}{4 n \left(\delta  \left(n+1\right) t^{1-n}+a_{0} \left(3n-1\right)\right)}
		\nonumber\\
		&\,+\frac{t^{3} \dot{F}}{2 n \left(a_{0} t^{n}+\delta  t\right) \left(n a_{0} t^{n}+\delta  t\right)}  . \nonumber 
	\end{align}
\normalsize

We will also find the Equations~\eqref{434a}--\eqref{434b} for various cases by applying the $F(t)=F_0\,t^r$ power-law ansatz.


\begin{thebibliography}{999}
		
		\bibitem{Krssak:2018ywd} {Krššák}, M.;  {van~den~Hoogen}, {R.J.};  {Pereira}, {J.G.};  {Boehmer}, {C.G.}; {Coley}, {A.A.} Teleparallel Theories of Gravity: Illuminating a Fully Invariant Approach. \textit{Class. Quantum Gravity} \textbf{2019}, \textit{{36}},  {183001}. arXiv:1810.12932.
		
		
		\bibitem{Krssak_Saridakis2015} {Krššák}, M.; {Saridakis}, {E.N.} The covariant formulation of $f(T)$ gravity. \textit{Class. Quantum Gravity} \textbf{2016}, \textit{33}, {115009}. arXiv:1510.08432.
		
		
		\bibitem{Coley:2019zld} Coley, {A.A.}; van den Hoogen, {R.J.}; McNutt, D.D. Symmetry and Equivalence in Teleparallel Gravity. \textit{J. Math. Phys.} \textbf{2020}, \textit{61}, 072503. arXiv:1911.03893.
		
		
		\bibitem{HJKP2018} Hohmann, M.;  Järv, L.;  Krššák, M.;  Pfeifer, C. Modified teleparallel theories of gravity in symmetric spacetimes. \textit{Phys. Rev. D} \textbf{2019}, \textit{100}, {084002}. arXiv:1901.05472.
		
		
		\bibitem{HJKP2018a} Hohmann, M.;  Järv, L.;  Krššák, M.;  Pfeifer, C. Teleparallel theories of gravity as analogue of non-linear electrodynamics. \textit{Phys. Rev.~D} \textbf{2018}, \textit{97}, 104042. arXiv:1711.09930.
		
		
		\bibitem{Bahamonde:2021gfp} Bahamonde, S.;  Dialektopoulos, K.F.;  Escamilla-Rivera, C.;  Farrugia, G.;  Gakis, V.;  Hendry, M.;  Hohmann, M.;  Said, J.L.;  Mifsud, J.; Di Valentino, E.  Teleparallel Gravity: From Theory to Cosmology. \textit{Rep. Prog. Phys.} \textbf{2023}, \textit{86}, 026901. arXiv:2106.13793.
		
		
		\bibitem{Cai_2015}	Cai, Y.-F.;  Capozziello, S.;  De Laurentis, M.; Saridakis, E.N. $f(T)$ teleparallel gravity and cosmology. \textit{Rep. Prog. Phys.} \textbf{2016}, \textit{79}, 106901. arXiv:1511.07586.
		
		
		\bibitem{preprint} Coley, A.A.;  van den Hoogen, R.J.; McNutt, D.D.  Symmetric Teleparallel Geometries. \textit{Class. Quantum Gravity} \textbf{2022}, \textit{39}, 22LT01. arXiv:2205.10719.
		
		
		
		\bibitem{SSpaper} Coley, A.A.;  Landry, A.;  van den Hoogen, R.J.; McNutt, D.D. Spherically symmetric teleparallel geometries. \textit{Eur. Phys. J. C} \textbf{2024}, \textit{84}, 334. arXiv:2402.07238.
		
		
		\bibitem{TdSpaper} Coley, A.A.;  Landry, A.;  van den Hoogen, R.J.; McNutt, D.D. Generalized Teleparallel de Sitter geometries. \textit{Eur. Phys. J. C} \textbf{2023}, \textit{83}, 977. arXiv:2307.12930. 
		
		
		\bibitem{coleylandrygholami} Coley, A.A.;  Landry, A.; Gholami, F. Teleparallel Robertson-Walker Geometries and Applications. \textit{Universe} \textbf{2023}, \textit{9}, 454. arXiv:2310.14378.
		
		
		\bibitem{nonvacSSpaper} Landry, A. Static spherically symmetric perfect fluid solutions in teleparallel F(T) gravity. \textit{Axioms} \textbf{2024}, \textit{13}, 333. arXiv:2405.09257.
		
		
		\bibitem{nonvacKSpaper} Landry, A. Kantowski-Sachs spherically symmetric solutions in teleparallel $F(T)$ gravity. \textit{Symmetry} \textbf{2024}, \textit{16}, 953. arXiv:2406.18659.
		
		
		\bibitem{roberthudsonSSpaper} van den Hoogen, R.J.; Forance, H. Teleparallel Geometry with Spherical Symmetry: The diagonal and proper frames. \textit{J. Cosmol. Astrophys.} \textbf{2024}, \emph{11}, 033. arXiv:2408.13342.
		
		
		\bibitem{Hohmann:2018rwf} Hohmann, M.; Järv, L.; Ualikhanova, U. Covariant formulation of scalar-torsion gravity. \textit{Phys. Rev. D} \textbf{2018} \textit{97}, {104011}. arXiv:1801.05786.
		
		
		\bibitem{Hohmann:2015pva} {Hohmann}, M. Spacetime and observer space symmetries in the language of Cartan geometry.  \textit{J. Math. Phys.} \textbf{2016}, \textit{57}, {082502}. arXiv:1505.07809.
		
		
		\bibitem{coley03} Coley, A.A. \emph{Dynamical Systems and Cosmology}; Kluwer Academic: Dordrecht, The Netherlands, 2003; ISBN 1-4020-1403-1.
		
		
		\bibitem{BahamondeBohmer} Bahamonde, S.;  Bohmer, C.G.; Carloni, S.;  Copeland, E.J.;  Fang, W.; Tamanini, N.  Dynamical systems applied to cosmology: Dark energy and modified gravity. \textit{Phys. Rep.} \textbf{2018}, \textit{775--777}, 1--122. arXiv:1712.03107.
		
		
		\bibitem{Kofinas} Kofinas, G.;  Leon, G.; Saridakis, E.N. Dynamical behavior in $f(T,T_G)$ cosmology. \textit{Class. Quantum Gravity} \textbf{2014}, \textit{31}, 175011. arXiv:1404.7100.
		
		
		\bibitem{BohmerJensko} Bohmer, C.G.; Jensko, E. Modified gravity: A unified approach to metric-affine models. \textit{J. Math. Phys.} \textbf{2023}, \textit{64}, 082505. arXiv:2301.11051.
		
		
		\bibitem{aldrovandi2003} Aldrovandi, R.;  Cuzinatto, R.R.; Medeiros, L.G. Analytic solutions for the $\Lambda$-FRW Model. \textit{Found. Phys.} \textbf{2006}, \textit{36}, 1736--1752. arXiv:gr-qc/0508073.
		
		
		\bibitem{bounce} Casalino, A.;  Sanna, B.;  Sebastiani, L.; Zerbini, S. Bounce Models within Teleparallel modified gravity.  \textit{Phys. Rev. D} \textbf{2021}, \textit{103}, {023514}. arXiv:2010.07609.
		
		
		\bibitem{Capozz} Capozziello, S.;  Luongo, O.;  Pincak, R.; Ravanpak, A. Cosmic acceleration in non-flat $f(T)$ cosmology. \textit{Gen. Relativ. Gravit.} \textbf{2018}, \textit{50}, 53. arXiv:1804.03649. 
		
		
		\bibitem{inflat} Bahamonde, S.;  Dialektopoulos, K.F.;  Hohmann, M.;  Said, J.L.;  Pfeifer, C.; Saridakis, E.N. Perturbations in Non-Flat Cosmology for $f(T)$ gravity. \textit{Eur. Phys. J. C} \textbf{2023}, \textit{83}, 193. arXiv:2203.00619.
		
		
		\bibitem{leonftbcosmo1} Paliathanasis, A.; Leon, G.  $f(T,B)$ gravity in a Friedmann--Lemaître--Robertson--Walker universe with nonzero spatial curvature. \textit{Math. Methods Appl. Sci.} \textbf{2023}, \textit{46}, 3905. arXiv:2201.12189.
		
		
		\bibitem{leonftbcosmo2} Paliathanasis, A.; Leon, G. Cosmological evolution in $f(T, B)$ gravity. \textit{Eur. Phys. J. Plus} \textbf{2021}, \textit{136}, 1092. arXiv:2106.01137.
		
		
		\bibitem{leon2} Leon, G.; Paliathanasis, A. Anisotropic spacetimes in $f(T,B)$ theory II: Kantowski-Sachs Universe. \textit{ Eur. Phys. J. Plus} \textbf{2022}, \textit{137}, 855. arXiv:2207.08570.
		
		
		\bibitem{leon3} Leon, G.; Paliathanasis, A. Anisotropic spacetimes in $f(T,B)$ theory III: LRS Bianchi III Universe. \textit{ Eur. Phys. J. Plus} \textbf{2022}, \textit{137}, 927. arXiv:2207.08571.
		
		
		\bibitem{cosmobaha} Bahamonde, S.; Zubair, M.; Abbas, G. Thermodynamics and cosmological reconstruction in $f(T,B)$ gravity. \textit{Phys. Dark Universe} \textbf{2018}, \textit{19}, 78. arXiv:1609.08373.
		
		
		\bibitem{dixit} Dixit, A.; Pradhan, A. Bulk Viscous Flat FLRW Model with Observational Constraints in $f(T,B)$ Gravity. \textit{Universe} \textbf{2022}, \textit{8}, 650.
		
		
		\bibitem{cosmoftbmodels} Zubair, M.;  Waheed, S.;  Fayyaz, M.A.; Ahmad, I. Energy Constraints and Phenomenon of Cosmic Evolution in $f(T,B)$ Framework. \textit{Eur. Phys. J. Plus} \textbf{2018}, \textit{133}, 452. arXiv:1807.07399.
		
		
		\bibitem{cosmoftbenergy} Bhattacharjee, S. Constraining $f(T,B)$ teleparallel gravity from energy conditions. \textit{New Astron.} \textbf{2021}, \textit{83}, 101495. arXiv:2004.12060.
		
		{
			\bibitem{ftbcosmo2} Caruana, M.;  Farrugia, G.; Said, J.L. Cosmological bouncing solutions in $f(T,B)$ gravity. \textit{Eur. Phys. J. C} \textbf{2020}, \textit{80}, 640. arXiv:2007.09925.
			
			\bibitem{ftbcosmo3} Chokyi, K.K.; Chattopadhyay, S. Cosmological Models within $f(T,B)$ Gravity in a Holographic Framework. \textit{Particles} \textbf{2024}, \textit{7}, 856.
			
			\bibitem{ftbcosmo4} Malik, A.;  Rauf, A.;  Venkatesha, V.;  Chalavadi, C.C.; Chaudhary, S. Dynamics of some cosmological solutions in modified $f(T,B)$ theory of gravity. \textit{Eur. Phys. J. Plus} \textbf{2024}, \textit{139}, 1008.
			
			\bibitem{ftbcosmo5} Escamilla-Rivera, C.; Said, L.J. Cosmological viable models in $f(T,B)$ gravity as solutions to the $H_0$ tension. \textit{Class. Quantum Gravity} \textbf{2020}, \textit{37}, 165002. arXiv:1909.10328.
			
			\bibitem{ftbcosmo6} Nájera, S.;  Aguilar, A.;  Rave-Franco, G.A.;  Escamilla-Rivera, C.; Sussman, R.A. Inhomogeneous solutions in $f(T,B)$ gravity. \textit{Int. J. Geom. Methods Mod. Phys.} \textbf{2022}, \textit{19}, 2240003. arXiv:2201.06177.	
		}
		
		\bibitem{ftphicosmo} Trivedi, O.;  Khlopov, M.;  Said, J.L.; Nunes, R. Cosmological singularities in $f(T,\phi)$ gravity. \textit{Eur. Phys. J. C} \textbf{2023}, \textit{83}, 1017. arXiv:2310.20222.
		
		
		\bibitem{ljsaidftb} Farrugia, G.;  Said, J.L.; Finch, A. Gravitoelectromagnetism, Solar System Test and Weak-Field Solutions in $f(T,B)$ Gravity with Observational Constraints. \textit{Universe} \textbf{2020}, \textit{6}, 34. arXiv:2002.08183. 
		
		
		{\bibitem{scalarfieldTRW} Landry, A. Scalar field sources Teleparallel Robertson-Walker $F(T)$-gravity {solutions.} \textit{Mathematics} \textbf{2024}, \emph{submitted}, \emph{Preprints}. \url{https://doi.org/10.20944/preprints202412.1896.v1}.} 
		
		\bibitem{scalarfieldKS} Landry, A. Scalar field Kantowski-Sachs spacetime solutions in teleparallel $F(T)$ {gravity.} {\textit{Universe} \textbf{2024}, \emph{submitted}, \emph{Preprints}. \url{https://doi.org/10.20944/preprints202412.1308.v1}.} 
		
		\bibitem{hypermomentum1} Iosifidis, D. Cosmological Hyperfluids, Torsion and Non-metricity. \textit{Eur. Phys. J. C} \textbf{2020}, \textit{80}, 1042. arXiv:2003.07384.
		
		
		\bibitem{hypermomentum2} Heisenberg, L.;  Hohmann, M.; Kuhn, S. Homogeneous and isotropic cosmology in general teleparallel gravity. \textit{Eur. Phys. J. C} \textbf{2023}, \textit{83}, 315. arXiv:2212.14324.
		
		\bibitem{hypermomentum3} Heisenberg, L.; Hohmann, M. Gauge-invariant cosmological perturbations in general teleparallel gravity. \textit{Eur. Phys. J. C} \textbf{2024}, \textit{84}, 462. arXiv:2311.05597.
		
		
		\bibitem{scalar1} Hohmann, M. Teleparallel gravity. In \textit{Modified and Quantum Gravity}; Springer, Cham., \textbf{2022}, \textit{1017}, {145.} arXiv:2207.06438. 
		
		
		\bibitem{scalar2} Hohmann, M. Scalar-torsion theories of gravity III: Analogue of scalar-tensor gravity and conformal invariants. \textit{Phys. Rev. D} \textbf{2018}, \textit{98}, 064004. arXiv:1801.06531.
		
		
		
		
		
		
		
		\bibitem{cosmofluidsbohmer} {Bohmer, C.G.;}
		d’Alfonso del Sordo, A. Cosmological fluids with boundary term couplings. \textit{Gen. Relativ. Gravit.} \textbf{2024}, \textit{56}, 75. arXiv:2404.05301. 
		
		
		\bibitem{steinhardt1} Zlatev, I.;  Wang, L.; Steinhardt, P. Quintessence, Cosmic Coincidence, and the Cosmological Constant. \textit{Phys. Rev. Lett.} \textbf{1999}, \textit{82}, 896. arXiv:astro-ph/9807002. 
		
		\bibitem{steinhardt2} Steinhardt, P.;  Wang, L.; Zlatev, I. Cosmological tracking solutions. \textit{Phys. Rev. D} \textbf{1999}, \textit{59}, 123504. arXiv:astro-ph/9812313. 
		
		
		\bibitem{caldwell1} Caldwell, R.R. A phantom menace? Cosmological consequences of a dark energy component with super-negative equation of state. \textit{Phys. Lett. B} \textbf{2002}, \textit{545}, 23. arXiv:astro-ph/9908168. 
		
		\bibitem{farnes} Farnes, J.S. A Unifying Theory of Dark Energy and Dark Matter: Negative Masses and Matter Creation within a Modified $\Lambda$CDM Framework. \textit{Astron. Astrophys.} \textbf{2018}, \textit{620}, A92. arXiv:1712.07962. 
		
		\bibitem{baumframpton} Baum, L.; Frampton, P.H. Turnaround in Cyclic Cosmology. \textit{Phys. Rev. Lett.} \textbf{2007}, \textit{98}, 071301. arXiv:hep-th/0610213. 
			
			{	\bibitem{quintom1} Cai, Y.-F.;  Saridakis, E.N.;  Setare, M.R.; Xia, J.-Q. Quintom Cosmology: Theoretical implications and observations. \textit{Phys. Rep.} \textbf{2010}, \textit{493}, 1. arXiv:0909.2776.
				
				\bibitem{quintom2} Guo, Z.-K.;  Piao, Y.-S.;  Zhang, X.; Zhang, Y.-Z. Cosmological evolution of a quintom model of dark energy. \textit{Phys. Lett. B} \textbf{2005}, \textit{608}, 177. arXiv:astro-ph/0410654.
				
				\bibitem{quintom3} Feng, B.;  Li, M.;  Piao, Y.-S.;  Zhang, X. Oscillating quintom and the recurrent universe. \textit{Phys. Lett. B} \textbf{2006}, \textit{634}, 101. arXiv:astro-ph/0407432.
				
				\bibitem{quintom4} Mishra, S.; Chakraborty, S. Dynamical system analysis of quintom dark energy model. \textit{Eur. Phys. J. C} \textbf{2018}, \textit{78}, 917. arXiv:1811.08279.
				
				\bibitem{quintomcoleytot} Tot, J.;  Coley, A.A.; Yildrim, B.; Leon, G. The dynamics of scalar-field Quintom cosmological models. \textit{Phys. Dark Universe} \textbf{2023}, \textit{39}, 101155. arXiv:2204.06538.
				
				\bibitem{quintomteleparallel1} Bahamonde, S.;  Marciu, M.; Rudra, P. Generalised teleparallel quintom dark energy non-minimally coupled with the scalar torsion and a boundary term. \textit{J. Cosmol. Astropart. Phys.} \textbf{2018}, \textit{04}, 056. arXiv:1802.09155.
			}
			
			
			
			
			
			
		\end{thebibliography}
\end{document}